\documentclass[preprint,showpacs,preprintnumbers,amsmath,amssymb]{revtex4}
\usepackage{graphicx,epsfig,dcolumn,bm,epic,eepic,float}
\usepackage{amsmath}
\usepackage{latexsym}
\usepackage{color}
\usepackage{cancel}
\usepackage{makeidx,shortvrb,latexsym}
\begin{document}
\unitlength 1 cm
\newcommand{\nn}{\nonumber}
\newcommand{\vk}{\vec k}
\newcommand{\vp}{\vec p}
\newcommand{\vq}{\vec q}
\newcommand{\vkp}{\vec {k'}}
\newcommand{\vpp}{\vec {p'}}
\newcommand{\vqp}{\vec {q'}}
\newcommand{\bk}{{\bf k}}
\newcommand{\bp}{{\bf p}}
\newcommand{\bq}{{\bf q}}
\newcommand{\br}{{\bf r}}
\newcommand{\bR}{{\bf R}}
\newcommand{\up}{\uparrow}
\newcommand{\down}{\downarrow}
\newcommand{\fns}{\footnotesize}
\newcommand{\ns}{\normalsize}
\newcommand{\cdag}{c^{\dagger}}
\title {The $NLO$  production of the $W^{\pm}$ and $Z^0$ vector
bosons via hadron collisions
 in the frameworks of $KMR$ and $MRW$
unintegrated parton distribution functions}
\author{$M. \; Modarres$ $^{a}$}
\altaffiliation {Corresponding author, Email: mmodares@ut.ac.ir,
Tel:+98-21-61118645, Fax:+98-21-88004781.}
\author{$M.R. \; Masouminia$ $^{a}$}
\altaffiliation {Visiting the Institute of Nuclear Physics, Polish
Academy of Science, Krakow, Poland.}
\author{$R. \; Aminzadeh\;Nik$ $^{a}$}
\author{$H. Hosseinkhani$$^{b}$}
\author{$N. Olanj$$^{c}$}
\affiliation {$^{a}$ Department of Physics, University of $Tehran$,
1439955961, $Tehran$, Iran.} \affiliation {$^{b}$Plasma and Fusion
Research School, Nuclear Science and Technology Research Institute,
14395-836, $Tehran$, Iran.} \affiliation {$^{c}$Physics Department,
Faculty of Science, $Bu$-$Ali Sina$ University, 65178, $Hamedan$,
Iran.}
\begin{abstract}
In a series of papers, we have investigated the compatibility of the
$Kimber$-$Martin$-$Ryskin$ ($KMR$) and $Martin$-$Ryskin$-$Watt$
($MRW$) $unintegrated$ parton distribution functions ($UPDF$) as
well as the  description of the experimental data on the proton
structure functions.  The present work is a sequel to that survey,
via calculation of the transverse momentum distribution of the
electro-weak gauge vector bosons in the $k_t$-factorization scheme,
by the means of the $KMR$, the $LO\;MRW$ and the $NLO\;MRW$ $UPDF$,
in the next-to leading order ($NLO$). To this end, we have
calculated and aggregated the invariant amplitudes of the
corresponding $involved$ diagrams in the $NLO$, and counted the
individual contributions in different frameworks. The preparation
process for the $UPDF$ utilizes the $PDF$ of $Martin$ et al,
$MSTW2008-LO$, $MSTW2008-NLO$, $MMHT2014-LO$ and $MMHT2014-NLO$ as
the inputs.  Afterwards, the results have been analyzed against each
other,  as well as  the existing experimental data. Our calculation
show excellent agreement with the experiment data. It is however
interesting to point-out that, the calculation using the $KMR$
framework illustrates a stronger agreement with the experimental
data, despite the fact that the $LO\;MRW$ and the $NLO\;MRW$
formalisms employ a better theoretical description of the $DGLAP$
evolution equation. This is of course due to the use of the
different implementation of the angular ordering constraint in the
$KMR$ approach, in which automatically includes the re-summation of
$ln({1/x})$, $BFKL$ logarithms, in the $LO$-$DGLAP$ evolution
equation.
\end{abstract}
 \maketitle
\section{Introduction}
In the recent years, new discoveries have been made at many high
energy particle physics laboratories, including the $LHC$,
concerning physics within the boundaries of the Standard Model and
beyond, as the consequence of pushing the maximum energy of the
experiments to the new limits. Today, many of these laboratories use
parton distribution functions ($PDF$) to describe and analysis their
extracted data from the deep inelastic $QCD$ collisions. These
scale-dependent functions are the solutions of the
$Dokshitzer$-$Gribov$-$Lipatov$-$Altarelli$-$Parisi$ ($DGLAP$)
evolution equations, \cite{DGLAP1,DGLAP2,DGLAP3,DGLAP4},
\begin{equation}
    {da(x,Q^2) \over dlog(Q^2)} =  {\alpha_s(Q^2) \over 2\pi}
    \sum_{b=q,g} \left[ \int_x^1 dz P_{ab}(z)b({x \over z},Q^2) - a(x,Q^2) \int_0^1 dz z P_{ba}(z)\right],
    \label{eq1}
\end{equation}
where $a(x,Q^2)$ can be either the distribution function of the
quarks, $xq(x,Q^2)$, or that of the gluons, $xg(x,Q^2)$, with $x$
being the fraction of the longitudinal momentum of the parent hadron
(the $Bjorken$ variable). The terms on the right-hand side of the
equation (\ref{eq1}), correspond to the real emission and the
virtual contributions, respectively. The scale $Q^2$ is an
ultra-violet cutoff, related to the virtuality of the exchanged
particle during the deep inelastic scattering ($DIS$). $P_{ab}(z)$
are the splitting functions of the respective partons which account
for the probability of emerging a parton $a(x'',Q^2)$ from a parent
parton $b(x',Q^2)$ through $z=x''/x'$.

The $DGLAP$ evolution equation however, is based on the
$strong\;ordering$ assumption, which systematically neglects the
transverse momentum of the emitted partons along the evolution
ladder. It has been repeatedly hinted that undermining the
contributions coming from the transverse momentum of the partons may
severely harm the precision of the calculations, especially in the
high energy processes in the small-$x$ region, see for example the
references \cite{KMR,MRW,KKMS,KIMBER,WattWZ}. This signaled the
necessity of introducing some transverse momentum dependent parton
distribution functions ($TMD\;PDF$), initially trough the
$Ciafaloni$-$Catani$-$Fiorani$-$Marchesini$ ($CCFM$) equation
\cite{CCFM1,CCFM2,CCFM3,CCFM4,CCFM5},
$$
    f(x,k_t^2,Q^2) = f_0(x,k_t^2,Q^2)
    + \int^{1}_{x} dz \int {dq^2 \over q^2} \Theta(Q-zq) \Delta_S(Q ,zq)
$$
\begin{equation}
    \times P(z,\bar\alpha_s(k_t^2))
    f({x \over z},|\mathbf{k_t} + (1-z)\mathbf{q}|^2,q^2).
    \label{eq2}
\end{equation}
The $\Theta(Q-zq)$ implies a physical condition, enforcing the
increase of the angle of the emission of the gluons in successive
radiations along the evolution chain. This condition which is
usually referred to as the angular ordering constraint ($AOC$), is
due to the coherent radiation of the gluons. The $Sudakov$ form
factor, $\Delta_S(Q,q)$, gives the probability of evolving from a
scale $q$ to a scale $Q$, without any partons emission, and can be
defined as:
\begin{equation}
    \Delta_S(Q ,q) = exp \left( -\bar{\alpha_s} \int^{Q^2}_{q^2} {dk^2 \over k^2}
    \int^{1}_{0} {dz^\prime} {1 \over (1-z)} \right) ,
    \label{eq3}
\end{equation}
with $\bar{\alpha_s}=3\alpha_s/\pi$. In the equation (\ref{eq2}),
$f(x,k_t^2,\mu^2)$ is the double-scaled $CCFM$ $TMD$ $PDF$, which in
addition to the $x$ and $Q$, depends on the transverse momentum of
the incoming partons, $k_t$. It has been shown (see the reference
\cite{CCFM-unfolding}) that in the proper boundaries, the $CCFM$
equation will reduced to the conventional $DGLAP$ and
$Balitski$-$Fadin$-$Kuraev$-$Lipatov$ ($BFKL$) equations,
\cite{BFKL1,BFKL2,BFKL3,BFKL4,BFKL5}.

The procedure of solving the $CCFM$ equation is mathematically
involved and unrealistically time consuming, since it includes
contemplating iterative integral equations with many terms. On the
other hand, the main feature of the $CCFM$ equation, i.e. the $AOC$,
can be exclusively used for the gluon evolution and therefore, this
process is incapable of producing convincing quark contribution. To
overcome these obstacles, Martin et al have introduced the
$k_t$-factorization framework and developed the
$Kimber$-$Martin$-$Ryskin$ ($KMR$) and the $Martin$-$Ryskin$-$Watt$
($MRW$) approaches \cite{KMR,MRW}, both of which are constructed
around the $LO$ $DGLAP$ evolution equations and modified with the
different visualizations of the angular ordering constraint. The
frameworks of $KMR$ and $MRW$ in the $LO$ and $NLO$ have been
investigated intensely in the recent years, see the references
\cite{Modarres1,Modarres2,Modarres3,Modarres4,Modarres5,Modarres6,Modarres7,Modarres8}.

Although $Martin$ et al have developed the $MRW$ formalism as an
improvement to the $KMR$ approach, by correcting the use of the
$AOC$, limiting its effect only on the diagonal splitting functions
and extending the range of their calculations into the $NLO$ via
introducing the $NLO\;MRW$ scheme, it appears that the $KMR$
approach, as an effective model, is more successful in producing a
realistic theory in order to describe the experiment. We are
therefore eager to expand our investigation regarding the merits and
shortcomings of these frameworks into the calculation of the
inclusive cross-sections of production of the electro-weak gauge
bosons in high energy hadronic collisions.

The process of the production of the massive gauge vector bosons,
$W^{\pm}$ and $Z^0$, have always been of extreme theoretical and
experimental interest, since it can provide invaluable information
about the nature of both the electro-weak and the strong
interactions, setting a benchmark for testing the validity of the
experiments and establishing a firm base for testing new theoretical
frameworks, see the references
\cite{UA1,UA2,CDF96,CDF2000,D095,D098,D02000-1,D02000-2,D02001,LIP1,Deak1}.
It is not however straightforward to describe the transverse
momentum distributions of the electro-weak bosons produced in
hadron-hadron collisions, since the usual collinear factorization
approach in the $LO$, neglects the transverse momentum dependency of
the incoming partons and therefore predicts a vanishing transverse
momentum for the product. Consequently, initial-state $QCD$
radiation is necessary to generate the $k_t$ distributions. On the
other hand, in this approximation, calculations for differential
cross sections of the $W^{\pm}$ and $Z^0$ production diverge
logarithmically in the $NLO$ limit for the $k_t \ll M_{W,Z}$ (which
is the main region of interest), due to the soft gluon emission. So,
one requires a re-summation to obtain a finite $k_t$ distribution.

In the present work we tend to calculate the $k_t$ distributions of
the cross-section of production of the $W^{\pm}$ and $Z^0$ using the
$NLO$ level diagrams and the $LO$ and $NLO$ $UPDF$ of the $KMR$ and
the $MRW$ frameworks. The $UPDF$ will be prepared in their proper
$k_t$-factorization schemes using the $PDF$ of $MSTW2008-LO$,
$MSTW2008-NLO$, $MMHT2014-LO$ and $MMHT2014-NLO$,
\cite{MSTW1,MSTW2,MSTW3,MMHT}. Such calculations have been
previously carried out using $LO$ matrix elements of quark-antiquark
annihilation cross section and doubly-unintegrated parton
distribution functions ($DUPDF$) in the framework of
$(k_t,z)$-factorization, reference \cite{WattWZ}, and in a
semi-$NLO$ approach, using a mixture of $LO$ and $NLO$ matrix
elements for the involved processes in addition to a variety of
$TMD$ $PDF$, see the reference \cite{LIP1}. To improve these
approximations and at the same time, test the functionality of the
$KMR$ and the $MRW$ $UPDF$, we have calculated the $NLO$ ladder
diagrams for $g + g \rightarrow W^{\pm}/Z^0 + q + q'$, $q + g
\rightarrow W^{\pm}/Z^0 + q' + g$ and $q + q' \rightarrow
W^{\pm}/Z^0 + g + g$, utilizing a physical gauge for the gluons. In
this way, at the price performing long and complicated calculations,
we will demonstrate that with the use of the $UPDF$ in the $NLO$
calculations, one can extract an excellent description of the
experimental data of the $D0$ [5,8,9] and $CDF$ [4] collaborations,
as well as others works given here, regarding the transverse
momentum distributions of the $W^{\pm}$ and $Z^0$ boson.

In what follows, first, a brief introduction to the concept of
$k_t$-factorization will be presented and the respective formalisms
for the $KMR$ and the $MRW$ frameworks will be derived, in the
section 2. The section 3 contains a comprehensive description over
the utilities and means for the calculation of the $k_t$-dependent
cross-section of production of the $W^{\pm}$ and $Z^0$ gauge vector
bosons in a hadron-hadron (or hadron-antihadron) deep inelastic
collision. The necessary numerical analysis will be presented in the
section 4, after which a thoroughgoing conclusion will be followed
 in the section 5.
\section{The $k_t$-factorization scheme}
A parton entering the sub-process at the top of the evolution
ladder, has non-negligible transverse momentum. However, it is
customary to use the $PDF$ of the $DGLAP$ or the $BFKL$ evolution
equations to describe such partons, despite the fact that these
density functions intrinsically carry no $k_t$-dependency. To
include the contributions coming from the transverse momentum
distributions of the partons, one can either use the solutions of
the $CCFM$ evolution equation or unify the $BFKL$ and the $DGLAP$
evolution equations to form a properly tuned $k_t$-dependent
framework, \cite{unified,stasto1}. Nevertheless, given the
mathematical complexity of these schemes, it is not desirable to use
them in the task of computing the $DIS$ cross-sections. Another way
is to convolute the single-scaled solutions of the $DGLAP$ evolution
equation and insert the required $k_t$-dependency via the process of
$k_t$-factorization (for a complete description see the reference
\cite{KIMBER}).

Thus, one may define the $UPDF$, $f_a(x,k_t^2,\mu^2)$, in the
$k_t$-factorization scheme, through the following normalization
relation,
\begin{equation}
    a(x,\mu^2) = \int^{\mu^2} {dk_t^2 \over k_t^2} f_a(x,k_t^2,\mu^2) ,\label{eq4}
    \end{equation}
where $a(x,\mu^2)$ are the solutions of the $DGLAP$ equation and
stand for either $xq(x,\mu^2)$ or $xg(x,\mu^2)$. The procedure of
deriving a direct expansion for $f_a(x,k_t^2,\mu^2)$, in terms of
the $PDF$ is strait forward. Yet, exposing the resulting
prescriptions to the different visualizations of the $AOC$ will
produce   different $UPDF$, namely the $KMR$, the $LO\;MRW$ and the
$NLO\;MRW$ frameworks. In what follows, we will describe these
frameworks in detail.
\subsection{The $KMR$ framework}
Starting form the $DGLAP$ equation in the leading order, the
equation (\ref{eq1}), and using the unregulated $LO$ $DGLAP$
splitting kernels, $P_{ab}(z)$,  the reference \cite{WATT}, $Kimber$
et al introduced an infrared cut-off, $\Delta$, as a visualization
of the $AOC$ \cite{KMR1},
    $$ \Theta(\theta - \theta') \Longrightarrow \mu > {zk_t \over (1-z)} \Longrightarrow \Delta = {k_t \over \mu + k_t}. $$
Limiting the upper boundary on $z$ integration by $\Delta$, excludes
$z=1$ form the integral equation and automatically prevents facing
the soft gluon singularities arising form the $1/(1-z)$ terms in the
splitting functions. Additionally, they factorized the virtual
contributions from the $DGLAP$ equations, by defining a virtual
(loop) contributions as:
    \begin{equation}
    T_a(k_t^2,\mu^2) = exp \left( - \int_{k_t^2}^{\mu^2} {\alpha_S(k^2) \over 2\pi}
    {dk^{2} \over k^2} \sum_{b=q,g} \int^{1-\Delta}_{0} dz' P_{ab}^{(LO)}(z') \right), \label{eq5}
    \end{equation}
    with
    $$ T_a(\mu^2,\mu^2) = 1, $$
as an appropriated form of the $Sudakov$ form factor, the equation
(\ref{eq3}). Afterwards, the double-scaled $KMR$ $UPDF$ are defined
as follows:
    \begin{equation}
    f_a(x,k_t^2,\mu^2) = T_a(k_t^2,\mu^2) \sum_{b=q,g} \left[ {\alpha_S(k_t^2) \over 2\pi}
    \int^{1-\Delta}_{x} dz P_{ab}^{(LO)}(z) b\left( {x \over z}, k_t^2 \right) \right] . \label{eq6}
    \end{equation}
According to the above formulation,  only at the last step of the
evolution, the dependence on the second scale, $\mu$, gets
introduced into the $UPDF$. The required $PDF$ is provided as input,
using the libraries $MSTW2008$ \cite{MSTW1,MSTW2,MSTW3} and
$MMHT2014$ \cite{MMHT}, where the calculation of the single-scaled
functions have been carried out using the $DIS$ data on the $F_2$
structure function of the proton. $T_a$ are  considered to be unity
for $k_t > \mu$. This constraint and its interpretation in terms of
the strong ordering condition gives the $KMR$ approach a smooth
behavior over the small-$x$ region, which is generally governed by
the $BFKL$ evolution equation.
\subsection{The $LO\;MRW$ framework}
In coordination with the theory of gluonic coherent radiation, it
has been pointed out that the $AOC$ in the $KMR$ formalism should
only act on the terms including the on-shell gluon emissions, i.e.
the diagonal splitting functions $P_{qq}(z)$ and $P_{gg}(z)$.
Therefore, $Martin$ et al defined the $LO$ $MRW$ $UPDF$ as the
correction to the $KMR$ framework \cite{MRW},
    $$
    f_q^{LO}(x,k_t^2,\mu^2)= T_q(k_t^2,\mu^2) {\alpha_S(k_t^2) \over 2\pi} \int_x^1 dz \left[
    P_{qq}^{(LO)}(z) {x \over z} q \left( {x \over z} , k_t^2 \right) \Theta \left( {\mu \over \mu + k_t}-z \right) \right.
    $$
    \begin{equation}
    \left. + P_{qg}^{(LO)}(z) {x \over z} g \left( {x \over z} , k_t^2 \right) \right], \label{eq7}
    \end{equation}
with
    \begin{equation}
    T_q(k_t^2,\mu^2) = exp \left( - \int_{k_t^2}^{\mu^2} {\alpha_S(k^2) \over 2\pi} {dk^{2} \over k^2}
    \int^{z_{max}}_{0} dz' P_{qq}^{(LO)}(z') \right), \label{eq8}
    \end{equation}
for the quarks and
    $$
    f_g^{LO}(x,k_t^2,\mu^2)= T_g(k_t^2,\mu^2) {\alpha_S(k_t^2) \over 2\pi} \int_x^1 dz \left[
    P_{gq}^{(LO)}(z)  \sum_q {x \over z} q \left( {x \over z} , k_t^2 \right)
    \right.$$
    \begin{equation}
    \left. + P_{gg}^{(LO)}(z) {x \over z} g \left( {x \over z} , k_t^2 \right) \Theta \left( {\mu \over \mu + k_t}-z \right)
    \right], \label{eq9}
    \end{equation}
with
    \begin{equation}
    T_g(k_t^2,\mu^2) = exp \left( - \int_{k_t^2}^{\mu^2} {\alpha_S(k^2) \over 2\pi} {dk^{2} \over k^2}
    \left[ \int^{z_{max}}_{z_{min}} dz' z' P_{qq}^{(LO)}(z')
    + n_f \int^1_0 dz' P_{qg}^{(LO)}(z') \right] \right) , \label{eq10}
    \end{equation}
for the gluons. In the equations (\ref{eq8}) and (\ref{eq10}),
$z_{max}=1-z_{min}=\mu/(\mu+k_t)$ \cite{WATT}. The $UPDF$ of $KMR$
and $MRW$ to a good approximation, include the main kinematical
effects involved in the $DIS$ processes. One should note that the
particular form of the $AOC$ in the $KMR$ formalism despite being of
the $LO$, includes some contributions from the $NLO$ sector, whence
in the case of $MRW$ framework, these contributions must be inserted
separately.
\subsection{The $NLO\;MRW$ framework}
The expansions of the $LO$ $MRW$ formalism into the $NLO$ region can
be achieved through the following definitions:
    $$
    f_a^{NLO}(x,k_t^2,\mu^2)= \int_x^1 dz T_a \left( k^2={k_t^2 \over (1-z)}, \mu^2 \right) {\alpha_S(k^2) \over 2\pi}
    \sum_{b=q,g} \tilde{P}_{ab}^{(LO+NLO)}(z)
    $$
    \begin{equation}
    \times b^{NLO} \left( {x \over z} , k^2 \right) \Theta \left( 1-z-{k_t^2 \over \mu^2} \right),
    \label{eq11}
    \end{equation}
with the $NLO$ splitting functions being defined as,
    \begin{equation}
    \tilde{P}_{ab}^{(LO+NLO)}(z) = \tilde{P}_{ab}^{(LO)}(z) + {\alpha_S \over 2\pi}
    \tilde{P}_{ab}^{(NLO)}(z),
    \label{eq12}
    \end{equation}
and
    \begin{equation}
    \tilde{P}_{ab}^{(i)}(z) = P_{ab}^{i}(z) - \Theta (z-(1-\Delta)) \delta_{ab} F^{i}_{ab} P_{ab}(z),
    \label{eq13}
    \end{equation}
where $i=0,1$ stand for $LO$ and $NLO$ respectively. The reader can
find a comprehensive description of the $NLO$ splitting functions in
the references \cite{MRW,PNLO}. We must however emphasis  that
contrary to the $KMR$ and the $LO$ $MRW$ frameworks, the $AOC$ is
being introduced into the $NLO$ $MRW$ formalism via the $\Theta
(z-(1-\Delta))$ constraint, in the "extended" splitting function.
Now $\Delta$ can be defined as:
    $$ \Delta = {k\sqrt{1-z} \over k\sqrt{1-z} + \mu}.$$
The $NLO$ corrections  introduced into this framework are  the
collection of the $NLO$ $PDF$, the $NLO$ splitting functions and the
constraint $\Theta \left( 1-z-{k_t^2 / \mu^2} \right)$.
Nevertheless, it has been shown that using only the $LO$ part of the
extended splitting function, instead of the complete definition of
equation (\ref{eq12}),  would result in reasonable accuracy in
computation of the $NLO$ $MRW$ $UPDF$ \cite{MRW}. Additionally, the
$Sudakov$ form factors in this framework are defined as:
    \begin{equation}
    T_q(k^2,\mu^2) = exp \left( - \int_{k^2}^{\mu^2} {\alpha_S(q^2) \over 2\pi} {dq^{2} \over q^2}
    \int^1_0 dz' z' \left[ \tilde{P}_{qq}^{(0+1)}(z') + \tilde{P}_{gq}^{(0+1)}(z') \right] \right) ,
    \label{eq14}
    \end{equation}
    \begin{equation}
    T_g(k^2,\mu^2) = exp \left( - \int_{k^2}^{\mu^2} {\alpha_S(q^2) \over 2\pi} {dq^{2} \over q^2}
    \int^1_0 dz' z' \left[ \tilde{P}_{gg}^{(0+1)}(z') + 2n_f\tilde{P}_{qg}^{(0+1)}(z') \right] \right) .
    \label{eq15}
    \end{equation}
Each of these $UPDF$, the $KMR$, $LO$ and $NLO$ $MRW$ can be used to
identify the probability of finding a parton of a given flavor, with
the fraction $x$ of longitudinal momentum of the parent hadron, the
transverse momentum $k_t$ in the scale $\mu$ at the semi-hard level
of a particular $DIS$ process. In the following section, we will
describe the cross-section of the production of the $W^{\pm}$ and
$Z^0$ bosons with the help of our $UPDF$.
\section{Production of $W^{\pm}$ and $Z^0$ in the $k_t$-factorization}
By definition, the total cross-section for a deep hadronic
collision, $\sigma_{Hadron-Hadron}$, can be written in terms of its
possible partonic constituents. Utilizing the $UPDF$ as density
functions for the involved partons, one may write
$\sigma_{Hadron-Hadron}$ in the following form:
    $$
    \sigma_{Hadron-Hadron} = \sum_{a_1,a_2=q,g} \int_0^1 {dx_1 \over x_1} \int_0^1 {dx_2 \over x_2}
    \int_{0}^{\infty} {dk^2_{1,t} \over k^2_{1,t}} \int_{0}^{\infty} {dk^2_{2,t} \over k^2_{2,t}}
    f_{a_1}(x_1,k^2_{1,t},\mu_1^2) f_{a_2}(x_2,k^2_{2,t},\mu_2^2)
    $$
    \begin{equation}
    \times
    \hat{\sigma}_{a_1 a_2}(x_1,k^2_{1,t},\mu_1^2;x_2,k^2_{2,t},\mu_2^2),
    \label{eq16}
    \end{equation}
where $a_1$ and $a_2$ are the incoming partons into the semi-hard
process from the first and the second hadrons, respectively.
$\hat{\sigma}_{a_1 a_2}$ are the corresponding partonic
cross-sections which can be defined separately as,
    \begin{equation}
    d\hat{\sigma}_{a_1 a_2} = {d\phi_{a_1 a_2} \over F_{a_1 a_2}} |{\mathcal{M}_{a_1 a_2}}|^2.
    \label{eq17}
    \end{equation}
$d\phi_{a_1 a_2}$ and $F_{a_1 a_2}$ are the multi-particle phase
space and the flux factor, respectively and can be defined according
to the specifications of the partonic process,
    \begin{equation}
    d\phi_{a_1 a_2} = \prod_i {d^3 p_i \over 2E_i} \delta^{(4)} \left( \sum p_{in} - \sum p_{out} \right) ,
    \label{eq18}
    \end{equation}
    \begin{equation}
    F_{a_1 a_2} = x_1 x_2 s,
    \label{eq19}
    \end{equation}
with the $s$ being the center of mass energy squared.
    $$ s=(P_1 + P_2)^2=2P_1.P_2. $$
$P_1$ and $P_2$ are the 4-momenta of the incoming protons and since
we are working in the infinite momentum frame, it is safe to neglect
their masses. $d\phi_{a_1 a_2}$ can be characterized in terms of
transverse momenta of the product particles, $p_{i,t}$, their
rapidities, $y_i$, and the azimuthal angles of the emissions,
$\varphi_i$,
    \begin{equation}
    {d^3 p_i \over 2E_i} = {\pi \over 2} dp_{i,t}^2 dy_i {d\varphi_i \over 2\pi}.
    \label{eq20}
    \end{equation}

In the equation (\ref{eq17}), ${\mathcal{M}_{a_1 a_2}}$ are the
matrix elements of the partonic diagrams which are involved in the
production of the final results. To calculate these quantities, one
must first understand the exact kinematics that rule over the
corresponding partonic processes.

\begin{figure}[ht]
\centering
\includegraphics[scale=0.2]{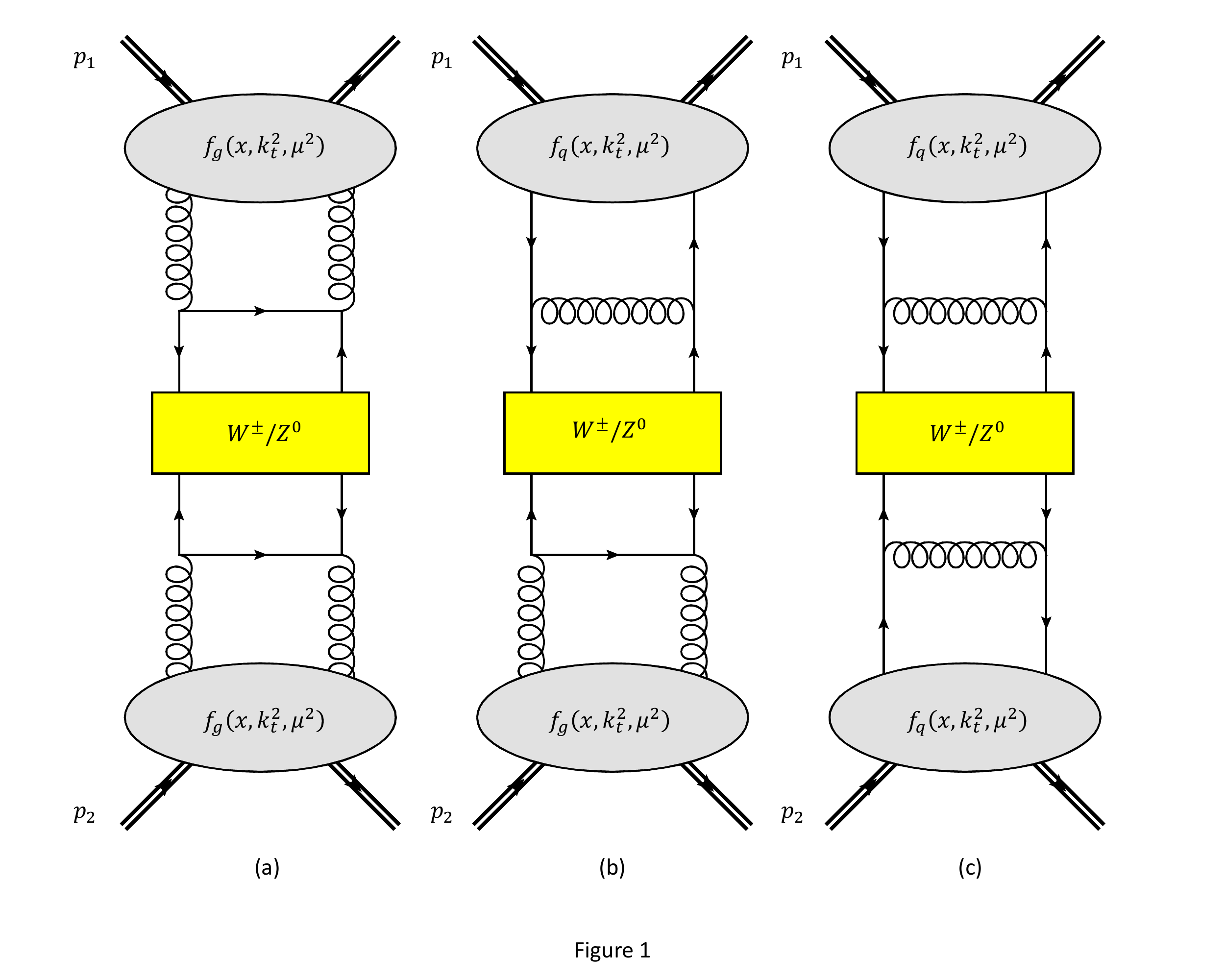}
\caption{The NLO ladder-type diagrams for the production of
$W^{\pm}$ and $Z^0$ in the $k_t$-factorization framework. The
$f_g(x,k_t^2,\mu^2)$ and $f_q(x,k_t^2,\mu^2)$ represent the
corresponding UPDF in the KMR, the  LO-MRW or the NLO-MRW
frameworks, i.e. the equations (\ref{eq6}), (\ref{eq7}), (\ref{eq9})
and (\ref{eq11}). } \label{fig1}
\end{figure}

The figure \ref{fig1} illustrates the ladder-type $NLO$ diagrams
that one have to consider, counting the contributions coming from $g
+ g \rightarrow W^{\pm}/Z^0 + q + q'$, $q + g \rightarrow
W^{\pm}/Z^0 + q' + g$, and $q + q' \rightarrow W^{\pm}/Z^0 + g + g$
as shown in the figure \ref{fig1}, panels (a), (b) and (c),
respectively. The kinematics and calculations of this type of
invariant amplitudes have been discussed extensively in the
references \cite{WattWZ,LIP1,Deak1}. We have followed the same
approach, obtaining the $dk_{i,t}^2/k_{i,t}^2$ terms only from the
ladder-type diagrams, and not from the interference (i.e. the
non-ladder) diagrams, using a physical gauge for the gluons, where
only the two transverse polarizations propagate,
    \begin{equation}
    d_{\mu \nu} (k) = -g_{\mu \nu} + { k_{\mu} n_{\nu} + n_{\mu} k_{\nu} \over k.n } .
    \label{eq21}
    \end{equation}
$n = x_1 P_1 + x_2 P_2$ is the gauge-fixing vector. Choosing such a
gauge condition, ensures that the $dk_{i,t}^2/k_{i,t}^2$ terms are
being obtained from the ladder-type diagrams on both sides of the
sub-processes. In the case of hadron-hadron collisions, one might
expect that neglecting the contributions coming from the non-ladder
diagrams, i.e. the diagrams where the production of the electro-weak
bosons is a by-product of the hadronic collision (see the reference
\cite{Deak1}), would have a numerical effect on the results.
Nevertheless, employing the gauge choice (\ref{eq21}), one finds out
that the contribution from the "unfactorizable" non-ladder diagrams
vanishes.

In the proton-antiproton center of mass frame, we can write the
following kinematics
    $$
    P_1 = {\sqrt{s} \over 2} (1,0,0,1), \;\;\; P_2 = {\sqrt{s} \over 2} (1,0,0,-1),
    $$
    \begin{equation}
    \textbf{k}_i = x_i \textbf{P}_i + \textbf{k}_{i,\perp}, \;\;\; k_{i,\perp}^2 = -k_{i,t}^2,  \;\;\; i=1,2 \; ,
    \label{eq22}
    \end{equation}
where the $k_i,\;i=1,2$ are the 4-momenta of the partons that enter
the semi-hard process. Afterwards, it is possible to write the law
of the transverse momentum conservation for the partonic process:
    \begin{equation}
    \textbf{k}_{1,\perp} + \textbf{k}_{2,\perp} = \textbf{p}_{1,\perp} + \textbf{p}_{2,\perp} + \textbf{p}_{\perp},
    \label{eq23}
    \end{equation}
with $\textbf{p}_{\perp}$ being the transverse momentum of the
produced vector boson. Additionally, defining the transverse mass of
the produced virtual partons, $m_{i,t}=\sqrt{m^2_i + p_{i}^2}$, we
can write,
    $$
    x_1 = \left( m_{1,t} e^{y_1} + m_{2,t} e^{y_2} + m_{W/Z,t} e^{y_{W/Z}} \right)/\sqrt{s},
    $$
    \begin{equation}
    x_2 = \left( m_{1,t} e^{-y_1} + m_{2,t} e^{-y_2} + m_{W/Z,t} e^{-y_{W/Z}} \right)/\sqrt{s}.
    \label{eq24}
    \end{equation}
Now, using the above equations, one can derive the following
equation for the total cross-section of the production of the
$W^{\pm}$ and $Z^0$ bosons in the framework of $k_t$-factorization,
    $$
    \sigma (P+\bar{P} \rightarrow W^{\pm}/Z^0 + X) = \sum_{a_i,b_i = q,g} \int
    {dk_{a_1,t}^2 \over k_{a_1,t}^2} \; {dk_{a_2,t}^2 \over k_{a_2,t}^2} \;
    dp_{b_1,t}^2 \; dp_{b_2,t}^2 \; dy_1 \; dy_2 \; dy_{W/Z}
    \; \times
    $$
    $$
    {d\varphi_{a_1} \over 2\pi} \; {d\varphi_{a_2} \over 2\pi} \; {d\varphi_{b_1} \over 2\pi} \; {d\varphi_{b_2} \over 2\pi}
    \times
    $$
    \begin{equation}
        {|\mathcal{M}(a_1+a_2 \rightarrow W^{\pm}/Z^0+b_1+b_2)|^2 \over 256 \pi^3 (x_1 x_2 s)^2} \;
        f_{a_1}(x_1,k_{a_1,t}^2,\mu^2) \; f_{a_2}(x_2,k_{a_2,t}^2,\mu^2).
    \label{eq25}
    \end{equation}
Note that the integration boundaries for ${dk_{i,t}^2 / k_{i,t}^2}$
are $(0,\infty)$. One may introduce an upper limit for these, say
$k_{i,max}$, several times larger than the scale $\mu$, without any
noticeable consequences. Yet, for $k_t<\mu_0$ with $\mu_0=1\;GeV$,
i.e. for the non-perturbative region, it is impervious to decide how
to validate our $UPDF$. A natural choice would be to fulfill the
requirement that
    $$
    \lim_{k_{a_i,t}^2 \rightarrow 0} f_{a_i}(x_i,k_{a_i,t}^2,\mu^2) \sim k_{a_i,t}^2,
    $$
and therefore, we can safely choose the following approximation for
the non-perturbative region:
    \begin{equation}
    f_{a_i}(x_i,k_{a_i,t}^2<\mu_0^2,\mu^2) = {k_{a_i,t}^2 \over \mu_0^2} a_i(x_i,\mu_0^2) T_{a_i}(\mu_0^2,\mu^2).
    \label{eq26}
    \end{equation}

In the next section, we will introduce some of the numerical methods
that have been used for the calculation of the $\sigma (P+\bar{P}
\rightarrow W^{\pm}/Z^0 + X)$, the equation (\ref{eq25}), using the
$UPDF$ of $KMR$ and $MRW$. It it expected that through considering
$NLO$ processes for this computation, the results will have a better
agreement with the existing experimental data, in comparison with
the previous calculations.
\section{Numerical analysis}
The main challenge one must face, in the computions of the total
cross-section of a hadron-hadron collision in the $NLO$, is the
extremely complex calculations required for extracting the invariant
amplitudes in a set of $2 \rightarrow 3$ $NLO$ Feynman diagrams.
Each of our processes, $g + g \rightarrow W^{\pm}/Z^0 + q + q'$, $q
+ g \rightarrow W^{\pm}/Z^0 + q' + g$, and $q + q' \rightarrow
W^{\pm}/Z^0 + g + g$, include a number of different configurations,
see  the  figure \ref{fig2}. This is when we filter out the
non-ladder diagrams, with our choice of the gauge condition on the
gluon polarization, the equation (\ref{eq21}). Writing the analytic
expressions of the $\mathcal{M}_{ab}$ for theses diagrams is rather
straight forward, see the Appendix A.

However, since the incoming and the out-going quarks are off-shell,
and we do not neglect their transverse momenta, their on-shell spin
density matrices has to be replaced with a more complicated
expression. To do this, one can extend the original expressions,
according to an approximation proposed in the references
\cite{LIP2,LIP3}, through converting the off-shell quark lines to
the internal lines via replacing the spinorial elements of the
incoming and the out-going partons. Following this idea, we replace
the incoming proton with a quark with the momentum $p$ and the mass
$m$ which radiates a photon or a gluon and turns into an off-shell
quark with the momentum $k$. Therefore, the corresponding matrix
element for such quarks can be written as,
    $$
    |\mathcal{M}|^2 \sim Tr \left( \Gamma_{\mu} \; {\hat{k}+m \over k^2 - m^2} \; \gamma^{\nu} \;
    u(p) \; \bar{u}(p)
    \; \gamma_{\nu} \; {\hat{k}+m \over k^2 - m^2} \; \Gamma^{\mu} \right)
    $$
where $\Gamma_{\mu}$ represents the rest of the original matrix
element. Now, the expression presented between $\Gamma_{\mu}$ and
$\Gamma^{\mu}$ is considered to be the off-shell quark spin density
matrix. Using the on-shell identity
    $$
    \sum u(p) \bar{u}(p) = \hat{p}+m,
    $$
and after performing some Dirac algebra at the $m \to 0$ limit, one
simply arrives to the following expression:
    $$
    |\mathcal{M}|^2 \sim {2 \over k^4} Tr \left( \Gamma_{\mu} \;
    \left[ k^2 \hat{p} - 2 (p \cdot k) \hat{k} \right] \;
    \Gamma^{\mu} \right).
    $$
Afterwards, imposing the Sudakov decomposition $k = xp + k_t$ with
$k^2 = k^2_t = -\textbf{k}^2_t$, one derives:
\begin{equation}
    |\mathcal{M}|^2 \sim {2 \over x k^2_{t}} \; Tr \left( \Gamma_{\mu} \;
    x \hat{p}
    \; \Gamma^{\mu} \right)\label{eq27}.
\end{equation}
Thus, with the above replacement, the negative light-cone momentum
fractions of the incoming partons have been neglected. $x \hat{p}$
in this equation represents the properly normalized off-shell spin
density matrix. Additionally, the coupling vertices of the off-shell
gluons to quarks must be modified with the eikonal vertex (i.e the
$BFKL$ prescription, see the reference \cite{Deak1}). Therefore, in
the case of initial off-shell gluons, we impose the so-called
non-sense polarization condition, i.e.
    $$
    \epsilon_{\mu}(k_i) = {2 k_{i,\mu} \over \sqrt{s}},
    $$
which results into the following normalization identity
    $$
    \sum \epsilon_{\mu}(k_i) \epsilon_{\nu}^{*}(k_i) = {k_{i,\mu} k_{i,\nu} \over k_{i,t}^2}.
    $$
We can calculate the evolution of the traces of the matrix elements
with the help of the algebraic manipulation system $\mathtt{FORM}$,
\cite{FORM}. Also, the method of orthogonal amplitudes, see the
reference \cite{Deak1}, can be used to further simplify the results.

The numerical computation of the equation (\ref{eq25}) have been
carried out using the $\mathtt{VEGAS}$ algorithm in the  Monte-Carlo
integration. To do this, we have selected the hard-scale of the
$UPDF$ to be equal to the transverse mass of the produced gauge
vector boson:
    $$
    \mu = (m_{W/Z}^2 + p_{W/Z,t}^2 )^{1 \over 2}.
    $$
Mathematically speaking, the upper bound on the transverse momentum
integrations of the master equation (25) should be the infinity.
However, since the $UPDF$ of $KMR$ and $MRW$ tend to quickly vanish
in the $k_t \gg \mu$ domain, one can safely introduce an ultraviolet
cut-off for these integrations. By convention, this cut-off is
considered to be at $k_{i,max} = p_{i,max} = 4 \mu$. Nevertheless,
given that $\mu$ depends on the transverse momentum of the produced
boson ($p_{W/Z,t}$) and its mass, it would be sufficient to set
$k_{i,max} = p_{i,max} = 4 \mu_{max}$, with
    $$
    \mu_{max} = (m_{W/Z}^2 + p_{t,max}^2 )^{1 \over 2}.
    $$
One can easily confirm that further domain have no contribution into
our results. Also it is satisfactory to bound the rapidity
integrations to $[-10,10]$, since $0 \leq x \leq 1$ and according to
the equation (\ref{eq24}), further domain has no contribution into
our results. The  choice of above hard scale is reasonable for the
production of W and Z bosons, as has been discussed in the reference
\cite{Deak1}.

As a final note, we should make it clear that in  the reference
\cite{LIP1}, the calculation of the transverse momentum distribution
for the production of the $W$ and $Z$ bosons has been carried out,
using the aggregated contributions of the following sub-processes:
\begin{itemize}
    \item[a)] The $NLO$ $g+g \to W/Z+q+\bar{q} $ partonic process, using the unintegrated gluon
    distributions of the $CCFM$ and the $LO$ $MRW$ formalisms, accounting for the production of the
    bosons accompanied by (at least) two distinct jets.
    \item[b)]The $LO$ $q+g \to W/Z+\bar{q} $ partonic process, with the density function of the
    incoming quarks and gluons being defined in the collinear ($GRV$ or $MSTW$) and the $k_t$-factorization
    (the $CCFM$ and the $LO$ $MRW$) formalisms, respectively. This corresponds to the $p+\bar{p} \to W/Z + jet + X$ cross-section.
    \item[c)]The $LO$ $q+\bar{q}  \to W/Z$ partonic process, from the collinear approximation,
    assuming that the incoming particles are valance quarks (or valance anti-quarks).
\end{itemize}
The above paratonic processes (a, b and c) obviously neglect some of
the $NLO$ contributions (in the b and c cases), namely the shares of
the non-valance quarks along the chain of evolution. Additionally,
assuming the non-zero transverse momentum for the valance quarks in
the infinite momentum frame is to some extent unacceptable, since,
in the absence of any extra structure, the intrinsic transverse
momenta of the valance quarks should not be enough for producing the
$W/Z$ bosons with relatively large $p_t$. In the present work, we
have upgraded the partonic processes of the b and c cases with their
NLO counterparts, i.e. $q^{*}+g^{*} \to W/Z+q +g$ and $q^{*}+\bar{q}
^{*} \to W/Z+g+g$ sub-processes. So, we are able to use the UPDF of
the $k_t$-factorization for the incoming quarks and gluons to insert
the transverse momentum dependency of the produced bosons, and at
the same time avoid over-counting. Furthermore, the problem of
separating the $W/Z$+single-jet and the $W/Z$+double-jet
cross-sections will reduce to inserting the correct physical
constraints on the dynamics of these processes, e.g. via inserting
some transverse momentum cuts for the produced jets, using the
anti-$k_t$ algorithm, see the reference \cite{anti-kt}.
Nevertheless, since we are interested to calculate the inclusive
cross-section for the production of the $W/Z$ bosons, inserting such
constraints is unnecessary.
\section{Results, Discussions and Conclusions}
Using the theory and the notions of the previous sections, one can
calculate  the production rate of the $W^{\pm}$ and $Z^0$ gauge
vector bosons for the center-of-mass energy of $1.8$ $TeV$. The
$PDF$ of Martin et al \cite{MSTW1,MSTW2,MSTW3,MMHT}, $MSTW2008$ and
$MMHT2014$, are used as the input functions to feed the equations
(\ref{eq6}), (\ref{eq7}), (\ref{eq9}) and (\ref{eq11}). The results
are the double-scale $UPDF$ in the $KMR$, the $LO \; MRW$ and the
$NLO \; MRW$ schemes. These $UPDF$ are in turn substituted into the
equation (\ref{eq25}) to construct the $W/Z$ cross-sections in their
respective frameworks. Since we intend to compare our calculations
to the $W^{\pm} \to l^{\pm} + \nu$ and $Z \to l^{+} + l^{-}$ decays,
we should multiply our theoretical out-put by the relevant branching
fractions, i.e. $f(W^{\pm} \to l^{\pm} + \nu) = 0.1075$ and $f(Z \to
l^{+} + l^{-}) = 0.033 66$ \cite{Yao}. Thus, the figures \ref{fig3}
 and \ref{fig4}
 present the reader,
with a comparison between the different contributions into the
differential cross-sections of the $W^{\pm}$ and $Z^0$, versus
 their transverse momentum ($k_t$) in the $KMR$ scheme. The
main contributions into the production of the $W^{\pm}$ are those
involving $u \to W + d$ and $c \to W + s$ vertices. Other production
vertices have been calculated and proven to be negligible compared
to these main contributions (nevertheless, for the sake of
completeness, we have included every single share, no matter how
small they are in the total contributions, see the figures
 \ref{fig5}
and
 \ref{fig6}, where the individual contributions of each of the
production vertices in the partonic sub-processes for the production
of $W^{\pm}$ and $Z^0$ have been depicted clearly, in the framework
of $KMR$ for $E_{CM}=1.8 \; TeV$). In the case of $Z^0$ production,
the main vertices are $u \to Z + u$, $d \to Z + d$, $c \to Z + c$
and $s \to Z + s$. In both cases, one can recognize the different
behavior of various partonic sub-processes. As expected, the
contributions of the $g+g \to W/Z + q + \bar{q}'$ in all of the
diagrams are similar, and even (roughly) of the same size, since
they only depend on the behavior of the gluon density. On the other
hand, the contribution coming from the $q+\bar{q}' \to W/Z + g + g$
differs from one production vertex to another, mimicking the
differences between the quark densities of different flavors and
going from the high contributions of the up and down quarks to small
contributions of the charm and strange and even negligible
contributions of the top and bottom quarks. Additionally, one
notices the smallness of the $q+g \to W/Z+q'+g$ contributions. This
is also anticipated, since the incoming gluon could (with a
relatively large probability) decay into a quark-antiquark pair that
does not have the right flavor to form a production vertex with
considerable contribution.

The figures \ref{fig7} and \ref{fig8}
 illustrate a complete comparison between the results of the
calculation of the production of the electro-weak gauge vector
bosons in the frameworks of $KMR$, $LO \; MRW$ and $NLO \; MRW$,
with each other and with the experimental data of the $D0$ and $CDF$
collaborations, references
\cite{CDF96,CDF2000,D098,D02000-1,D02000-2,D02001}. The results in
the $KMR$ framework has an excellent agreement with the experimental
data, both in the $W^{\pm}$ and $Z^0$ productions. The $LO \; MRW$
scheme behaves similarly compared to the $KMR$ framework, yet has a
noticeably shorter peak, specially in the case of $Z^0$. This is due
the different visualization of the $AOC$ between these two
frameworks, see the section 3. Meanwhile, the results in the $NLO \;
MRW$ scheme are unexpectedly unable to describe the experiment data.
This is related to the conditions in which the $AOC$ has been
imposed in this framework. The $\theta(1-z-{k_t^2 / \mu^2})$
constraint gives the parton distributions of the $NLO \; MRW$, a
sharp descend to zero at $k_t \to \mu$ and returns a vanishing
contribution for the better part of the transverse momentum
integration in the equation (\ref{eq25}). Consequently, the overall
value of the differential cross-sections of the $W^{\pm}$ and $Z^0$
production in this framework reduces dramatically, as it is apparent
in the figures \ref{fig7} and  \ref{fig8} . In overall and as it has
been stated elsewhere (see for example the references
\cite{Modarres7, Modarres8}) the results in the $KMR$ scheme
seemingly have a better agreement with the experiment. This is to
some extend ironic, since the $LO$ and the $NLO \; MRW$ formalisms
are developed as extensions and improvements to the $KMR$ approach
and are more compatible with the $DGLAP$ evolution equation.

Such comparisons can also be made for the larger values of $k_t$,
see the  figures
 \ref{fig9} and \ref{fig10}, where the production rates of the electro-weak gauge bosons are
plotted against their transverse momentum for $k_t < 200 \; GeV$.
The diagrams include the calculations of $d\sigma_{W/Z}/dk_t$ and
$1/\sigma_{W} \; d\sigma_{W}/dk_t$ and the comparisons are made with
the help of the data from the $D0$ collaboration, references
\cite{D098,D02001}. Of course, since the data points have small
values and large errors, and because of the closeness of the results
in different frameworks, one cannot stress over the superiority of
any of the approaches. Yet, our previous conclusion about the
validity of the $KMR$ $UPDF$ and the short-comings of the $NLO \;
MRW$ $UPDF$ holds. Another interesting observation is that in the
large $k_t$, where because of the smallness of the results the
higher order corrections become important, the calculations in the
$KMR$ approach start to separate from the $LO \; MRW$ and behave
similar to the $NLO \; MRW$. The reason is that the inclusion of the
$non-diagonal$ splitting functions into the domain of the $AOC$
introduces some corrections from the $NLO$ region. Additionally, one
notices that the contribution coming from the $q+q' \to W^{\pm} + g
+ g$ in the $NLO$ evaluations considerably deviates from the similar
behavior of its respective counterparts. This of course roots in the
evolution of the $NLO$ quark densities in this framework, see the
reference \cite{WATT}

Recently, Martin et al have updated their $PDF$ libraries, the
reference \cite{MMHT}. The figures
 \ref{fig11} and \ref{fig12}
demonstrate the differences between the cross-section of the
production of the $W/Z$ vector bosons in the $KMR$ framework, using
the (older) $MSTW2008$ and the (newer) $MMHT2014$ $PDF$. One notices
that,  using either of these $PDF$ as input for our $UPDF$ produces
a negligible difference.

The figures \ref{fig13} and \ref{fig14} present an interesting
comparison between the experimental data and the results of the
different approximations in the calculation of the production of the
electro-weak gauge vector bosons. In addition to our calculations in
the $KMR$ and the $MRW$ $UPDF$ in the $LO$ and the $NLO$
approximations, the results coming from  the $CCFM$ $TMD$ $PDF$
(reference \cite{LIP1}), the $doublly$ $unintegrated$ parton
distributions ($DUPDF$, see the reference \cite{WattWZ}) and from
the $collinear$ frameworks are included in these diagrams. The
$CCFM$ results are calculated as the sum of $g+g \to W/Z + q +
\bar{q}'$, $g+q \to W/Z + q'$ and $q+q \to W/Z$ sub-processes. The
$DUPDF$ results are in the $(k_t-z)$-factorization framework,
utilizing a $q+q \to W/Z$ "effective" production vertex.
Furthermore, to calculate the differential cross-section of the
$W/Z$ production in the $collinear$ approximation, one have to
ignore the transverse momentum integrations in the equation
(\ref{eq25}) and replace the $UPDF$ with the unpolarized parton
distributions of $MSTW2008$, $MMHT2014$ or $GRV2009$
\cite{GRV1,GRV2,GRV3}:
    $$
    \sigma (P+\bar{P} \rightarrow W^{\pm}/Z^0 + X) = \sum_{a_i,b_i = q,g} \int
    dp_{b_1,t}^2 \; dp_{b_2,t}^2 \; dy_1 \; dy_2 \; dy_{W/Z} \;
    {d\varphi_{b_1} \over 2\pi} \; {d\varphi_{b_2} \over 2\pi}
    \times
    $$
    \begin{equation}
        {|\mathcal{M}(a_1+a_2 \rightarrow W/Z+b_1+b_2)|^2 \over 256 \pi^3 (x_1 x_2 s)^2} \;
        a_1(x_1,\mu^2) \; a_2(x_2,\mu^2).
    \label{eq28}
    \end{equation}

The reader should notice that the results of our computations in the
$NLO$ regime, as expected, have a better behavior towards describing
the experimental data, both in the $W^{\pm}$ and $Z^0$ cases, since
they descend with a shallow steep, compared to the results
calculated in other schemes. This is in part, because the $NLO$
evaluations are inherently more accurate. Yet, most of the credit
goes to the precision of the utilized $UPDF$. Again, the $KMR$
framework in the $NLO$ calculations offers the best description of
the experiment.

Additionally, it is possible to compare our presumed frameworks
through the calculation of the total cross-section of the $W^{\pm}$
and $Z^0$ production with respect to the center-of-mass energy of
the hadronic collision, i.e. the  figures
 \ref{fig15} and \ref{fig16}.
Following our previous pattern, the results of both the $KMR$ and
the $LO \; MRW$ frameworks show a good level of compatibility with
the experimental data. On the other hand, since the $NLO \; MRW$
framework has failed to describe the data, we have excluded its
contributions here, to save some computation time.

Finally, it has been brought to our attention that the $ATLAS$ and
$CMS$ collaborations have recently published some data regarding the
production of the $Z^0$ gauge vector boson in the $LHC$ for $E_{CM}
= 8\;TeV$, the references \cite{ATLAS2016,CMS2015}. In the above
calculations, the rapidity of the produced boson has been separated
in equally spaced rapidity sectors within $0 <|y_Z| <2.4$ domain. In
figure \ref{fig17}, we have addressed the above observations, using
our $NLO$ framework and utilizing the $UPDF$ of $KMR$, since we have
already established the superiority of this scheme in describing the
experiment. The individual contributions from the partonic
sub-processes are presented and the total values of (single and
double) differential cross-sections are subjected to comparison with
the data of the $ATLAS$ and $CMS$ collaborations. One easily notes
that our calculations is in general agreement with the experimental
data and with similar calculations in a $NNLO$ $QCD$ framework from
the reference \cite{NNLOQCD-Z}.

Unfortunately, performing these calculations are extremely
time-consuming and the existing data points are not plentiful or
accurate enough to let us make a decisive statement about the
superiority regarding any of our presumed frameworks. Nevertheless,
considering these comparisons, it is apparent that the $KMR$ $UPDF$
in the framework of $k_t$-factorization, despite their
miss-alignments with the theory of the $DGLAP$ evolution equation
and the physics of the successive gluon radiations, as an effective
theory, proposes the best option to describe the deep inelastic
$QCD$ events. However, until further phenomenological analysis, such
claim remains as an educated speculation.

In summary, within the present work, we have calculated the rate of
productions belonging to the electro-weak gauge vector bosons in the
framework of $k_t$-factorization, utilizing the $UPDF$ of $KMR$, $LO
\; MRW$ and $NLO \; MRW$, by the means of $NLO$ $QCD$ processes. The
results have been demonstrated and compared to each other and to the
experimental data points from the $D0$ and the $CDF$ collaborations,
as well as the calculations in other frameworks. Through our
analysis we have suggested that despite the theoretical advantages
of the $MRW$ formalism, the $KMR$ approach has a better behavior
toward describing the experiment.

\begin{acknowledgements}
$MM$ would like to acknowledge the Research Council of University of
Tehran and the Institute for Research and Planning in Higher
Education for the grants provided for him.

$MRM$ sincerely thanks A. Lipatov and N. Darvishi for their valuable
discussions and comments. $MRM$ extends his gratitude towards his
kind hosts at the Institute of Nuclear Physics, Polish Academy of
Science for their hospitality during his visit. He also acknowledges
the Ministry of Science, Research and Technology of Iran that funded
his visit.
\end{acknowledgements}

\appendix
\section{The matrix elements of the partonic sub-processes}
\begin{figure}[ht]
\centering
\includegraphics[scale=0.2]{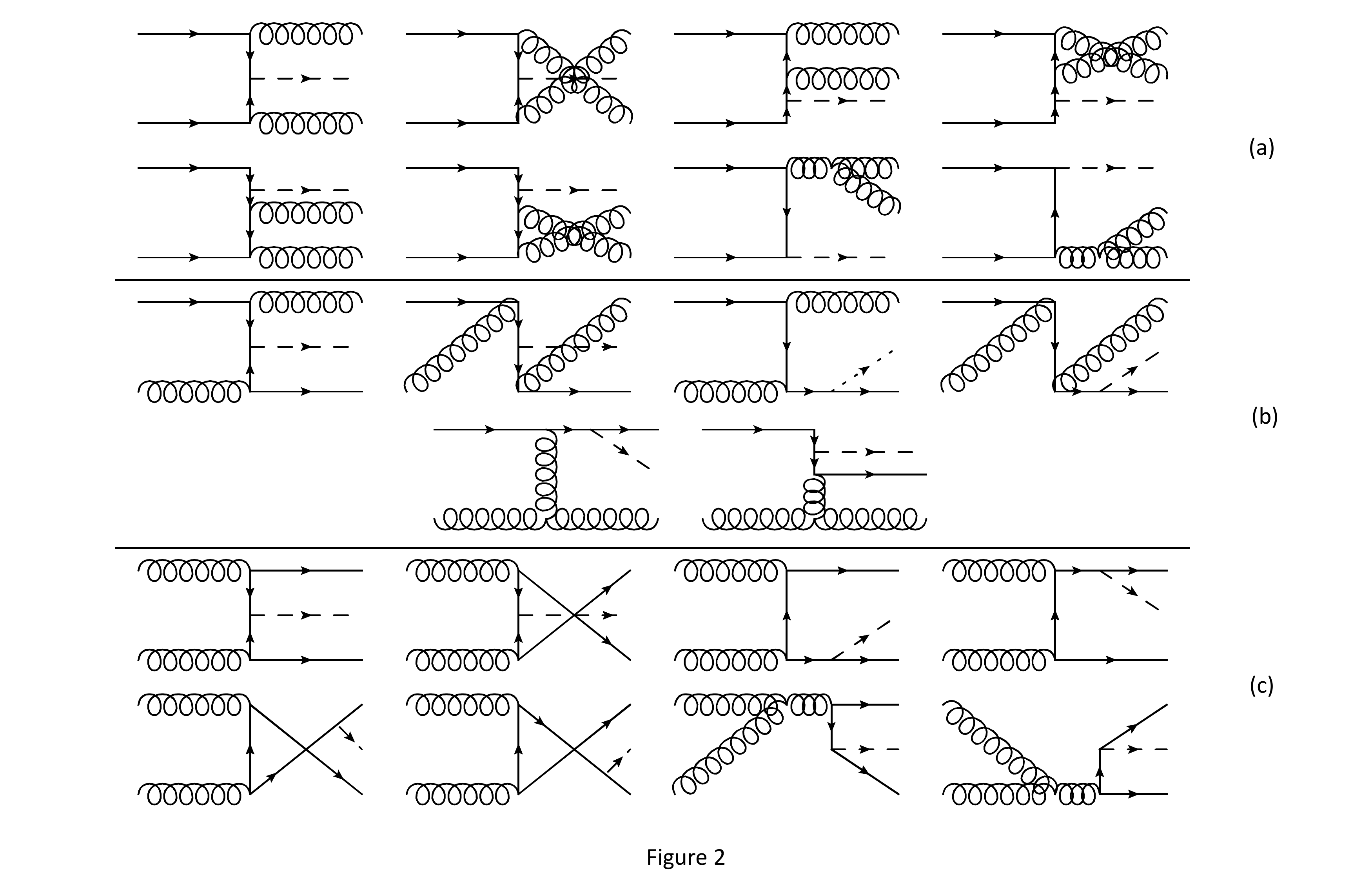}
\caption{The individual contributions into the matrix elements of
the partonic scattering. The diagrams in the panel (a) correspond to
the $q+\bar{q}' \rightarrow W^{\pm}/Z^0 + g + g$ sub-process, panel
(b) to the $q + g  \rightarrow W^{\pm}/Z^0 + q + g$ sub-process and
panel (c) to the $g+g \rightarrow W^{\pm}/Z^0 + q + \bar{q}'$
sub-process. It should be pointed out  that one may find additional
non-ladder-type diagrams which contribute to these matrix elements.
We have eliminated these un-desirable contributions using our choice
of the gluon gauge, the equation (\ref{eq21}).} \label{fig2}
\end{figure}

\begin{figure}[ht]
\centering
\includegraphics[scale=0.2]{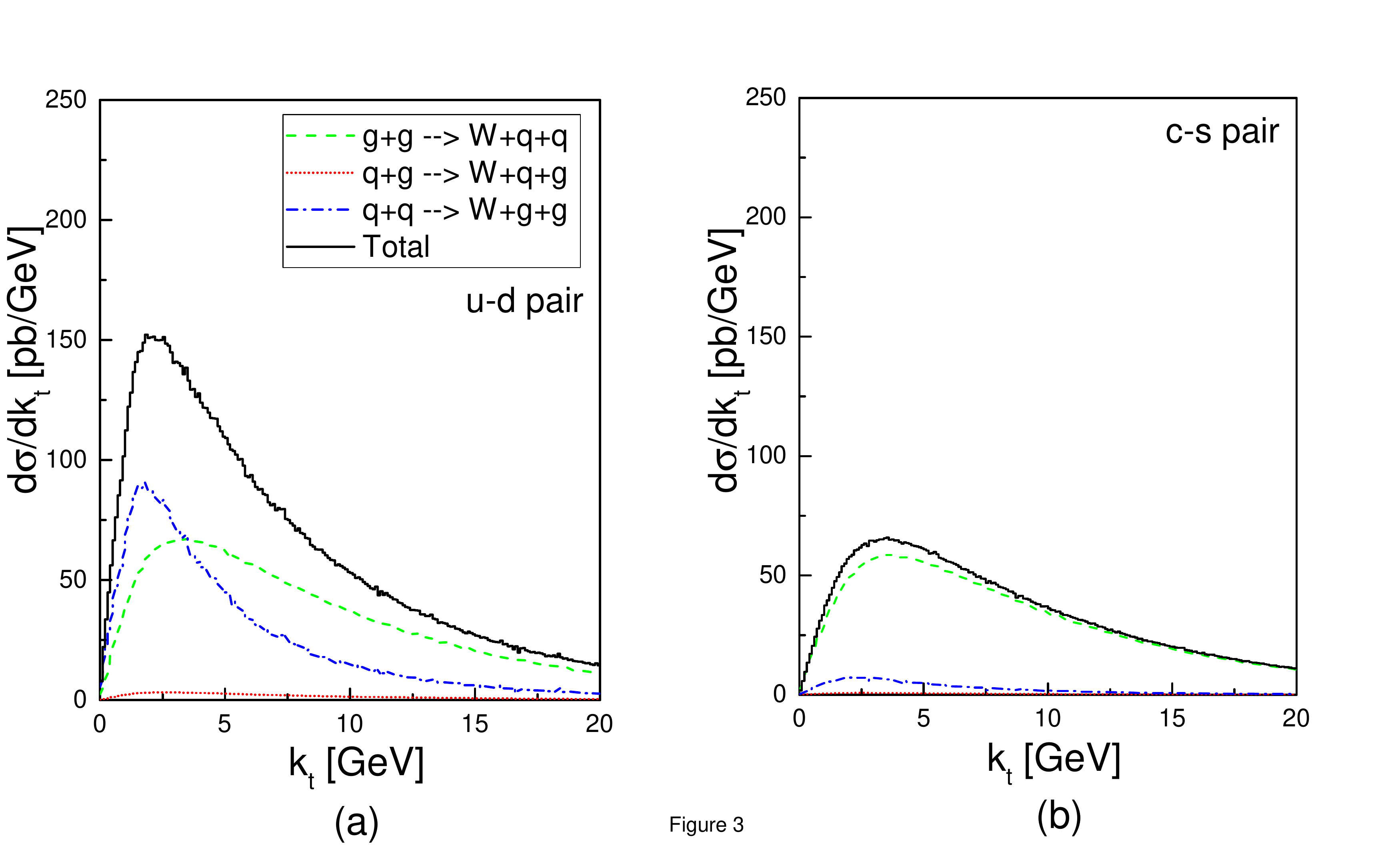}
\caption{The differential cross-section of the productions of
$W^{\pm}$ bosons in a DIS at $E_{CM}=1.8\;TeV$, against the
transverse momentum distribution of the produced particle. The
panels (a) and (b) illustrate the up-down and charm-strange
contributions, respectively. The contribution of each partonic
sub-process is singled out: the green-dash histogram is for $g+g
\rightarrow W^{\pm} + q + \bar{q}'$, the red-dotted histogram is for
$q + g \rightarrow W^{\pm} + q' + g$ and the blue-dash-dotted
histogram  is for $q+\bar{q}' \rightarrow W^{\pm} + g + g$. The
black-full histogram is the total contribution of the give quark
pairs. The histograms are produced using the $KMR$ $UPDF$ with the
$PDF$ of $MSTW2008$. } \label{fig3}
\end{figure}

\begin{figure}[ht]
\centering
\includegraphics[scale=0.2]{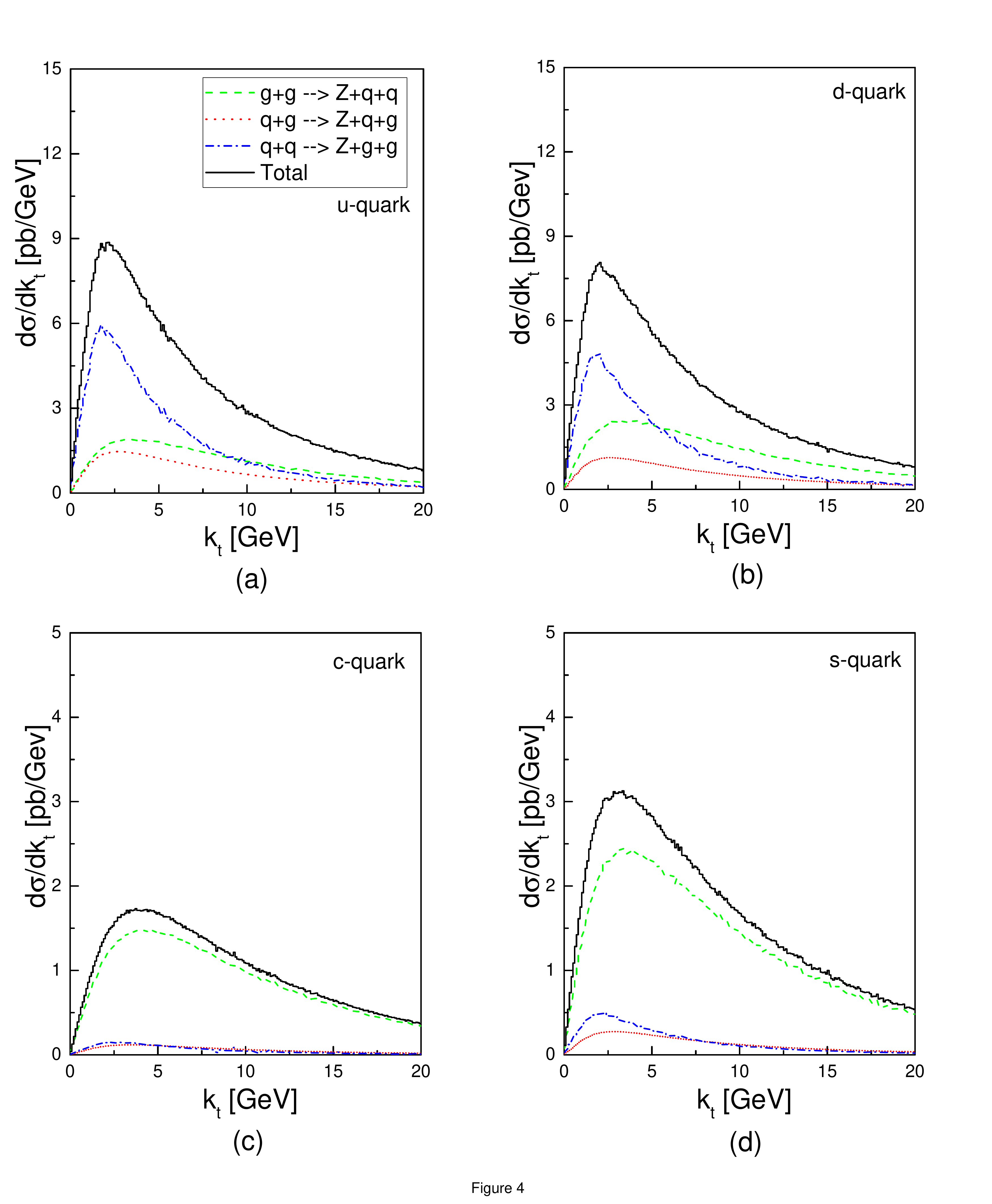}
\caption{The differential cross-section of the productions of
$Z^{0}$ boson in a DIS at $E_{CM}=1.8\;TeV$, against the transverse
momentum distribution of the produced particle. The contributions of
the up and the down quarks (the panels (a) and (b), respectively)
and the lightest sea-quarks (the panel (c) for the charm quark and
the panel (d) for the strange quark). The green-dash histogram is
for $g+g \rightarrow Z^{0} + q + \bar{q}$, the red-dotted histogram
is for $q + g \rightarrow Z^{0} + q + g$ and the blue-dash-dotted
histogram is for $q+\bar{q} \rightarrow Z^{0} + g + g$. The
black-full histogram is the total contribution of the given quark.
The data is produced using the $KMR$ $UPDF$, with the $PDF$ of
$MSTW2008$.} \label{fig4}
\end{figure}

\begin{figure}[ht]
\centering
\includegraphics[scale=0.2]{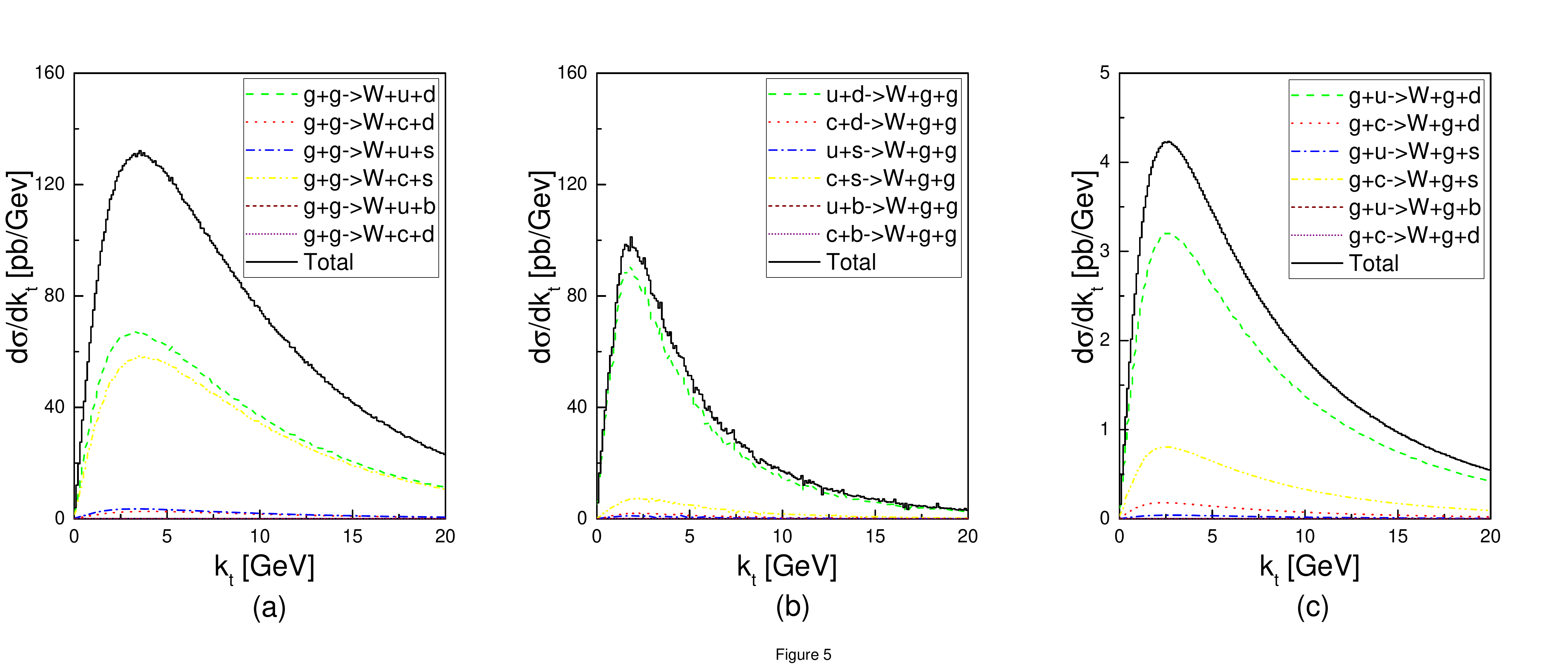}
\caption{The contributions of the individual partonic sub-processes
into the differential cross-section of the productions of $W^{\pm}$
bosons in a DIS at $E_{CM}=1.8\;TeV$, versus the transverse momentum
distribution of the produced particle. The panels (a), (b) and (c)
correspond to the $g+g \rightarrow W^{\pm} + q + \bar{q}'$,
$q+\bar{q}' \rightarrow W^{\pm} + g+g$ and $g+q \rightarrow W^{\pm}
+ g + q'$ sub-processes, respectively. The data have been obtained
using the $UPDF$ of $KMR$, with  the $PDF$ of $MSTW2008$.}
\label{fig5}
\end{figure}

\begin{figure}[ht]
\centering
\includegraphics[scale=0.2]{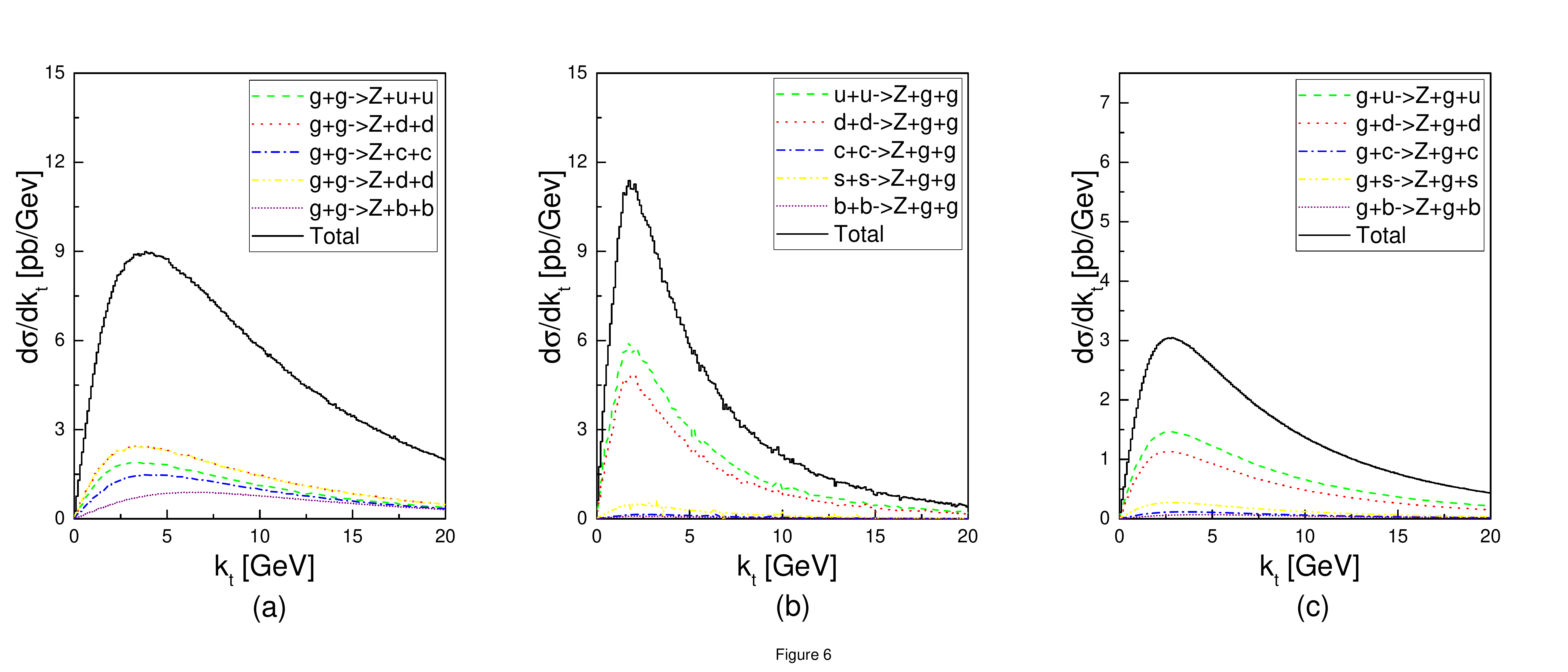}
\caption{The contributions of the individual partonic sub-processes
into the differential cross-section of the productions of $Z^{0}$
bosons in a DIS at $E_{CM}=1.8\;TeV$, versus the transverse momentum
distribution of the produced particle. The notions of the diagrams
are the same as in the figure \ref{fig5}.} \label{fig6}
\end{figure}

\begin{figure}[ht]
\centering
\includegraphics[scale=0.2]{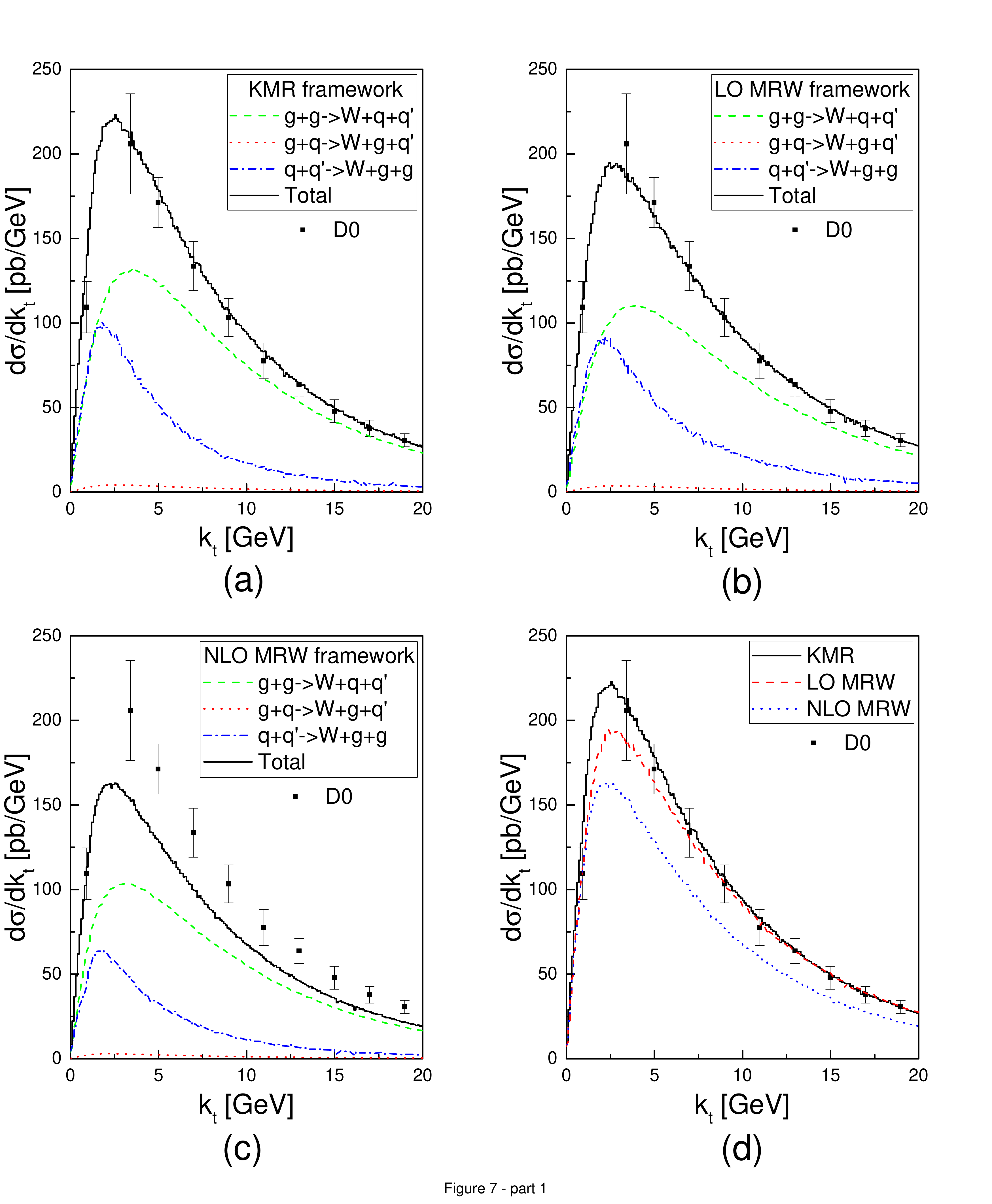}
\includegraphics[scale=0.2]{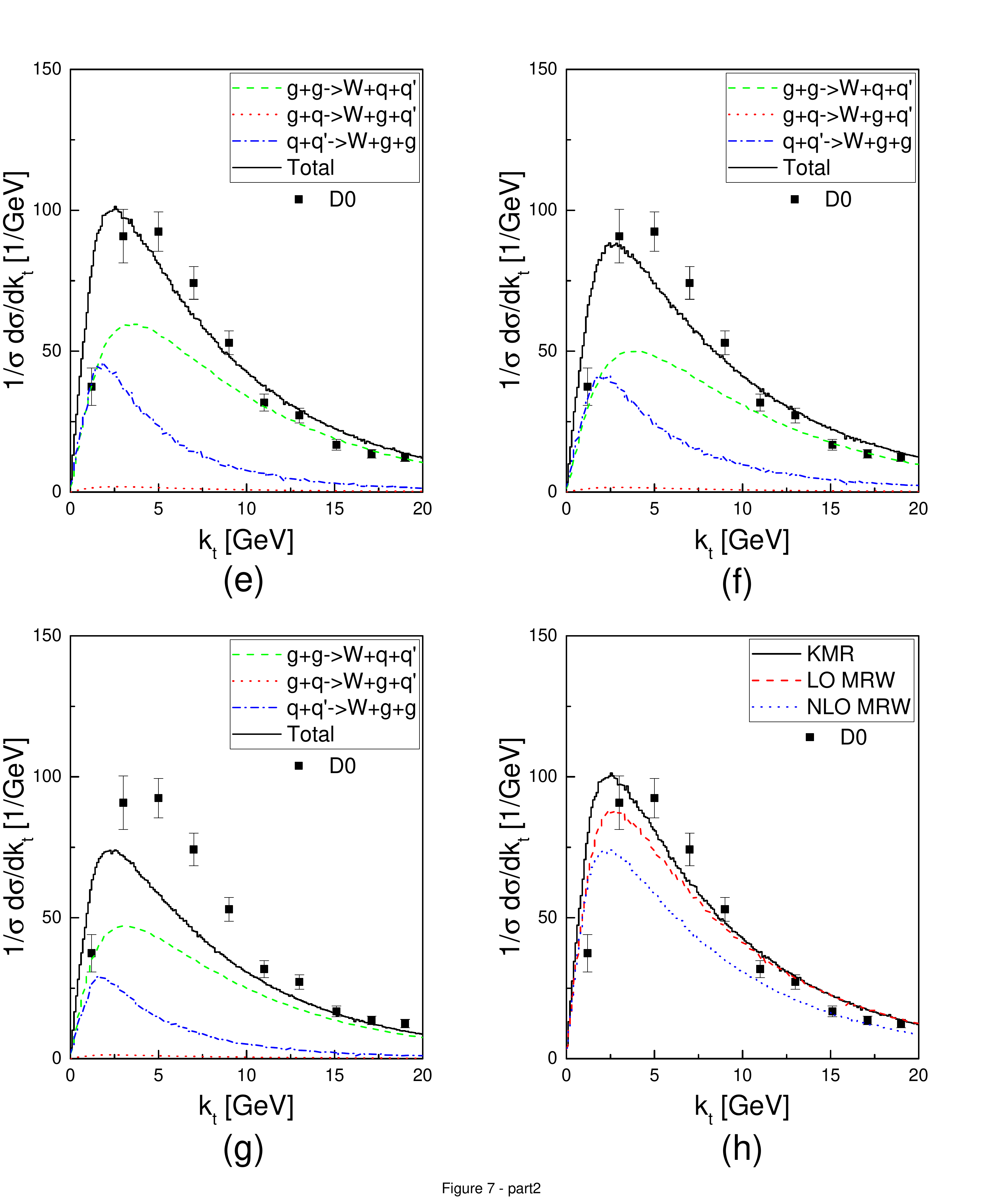}
\caption{The comparison of the differential cross-section of the
$W^{\pm}$ production in the $NLO$ in the $KMR$ (the panel (a)),
$LO\;MRW$ (the panel (b)) and $NLO\;MRW$ (the panel (c)) frameworks.
The panel (d) illustrates this comparison with the help of
experimental data of $D0$ collaboration, the reference
\cite{D02001}. The panels (e),(f) and (g) are the same values, but
this time they deviled by the total cross-sections in their
respective framework and compared to an older set of data points
from the $D0$ collaboration, \cite{D098}. Again, an overall
comparison with the experiment is presented in the panel (h). To
perform these calculations, we have utilized the $PDF$ of
$MSTW2008$.} \label{fig7}
\end{figure}

\begin{figure}[ht]
\centering
\includegraphics[scale=0.2]{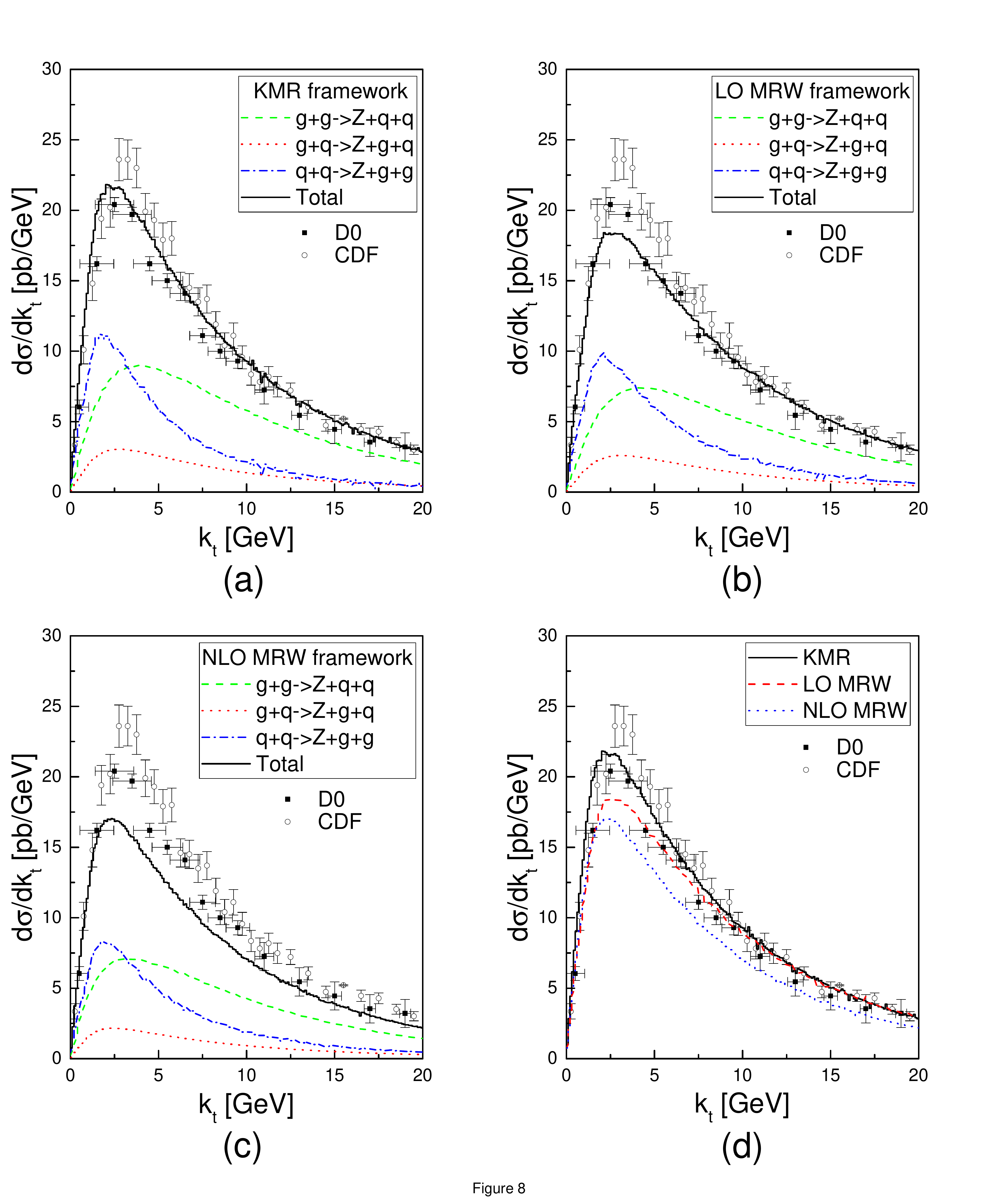}
\caption{The comparison of the differential cross-section of the
$Z^{0}$ production in the $NLO$ in the $KMR$ (the panel (a)),
$LO\;MRW$ (the panel (b)) and $NLO\;MRW$ (the panel (c)) frameworks.
The panel (d) illustrates this comparison with the help of the
experimental data of $D0$ and $CDF$ collaborations, the references
\cite{D098,CDF2000}.} \label{fig8}
\end{figure}

\begin{figure}[ht]
\centering
\includegraphics[scale=0.2]{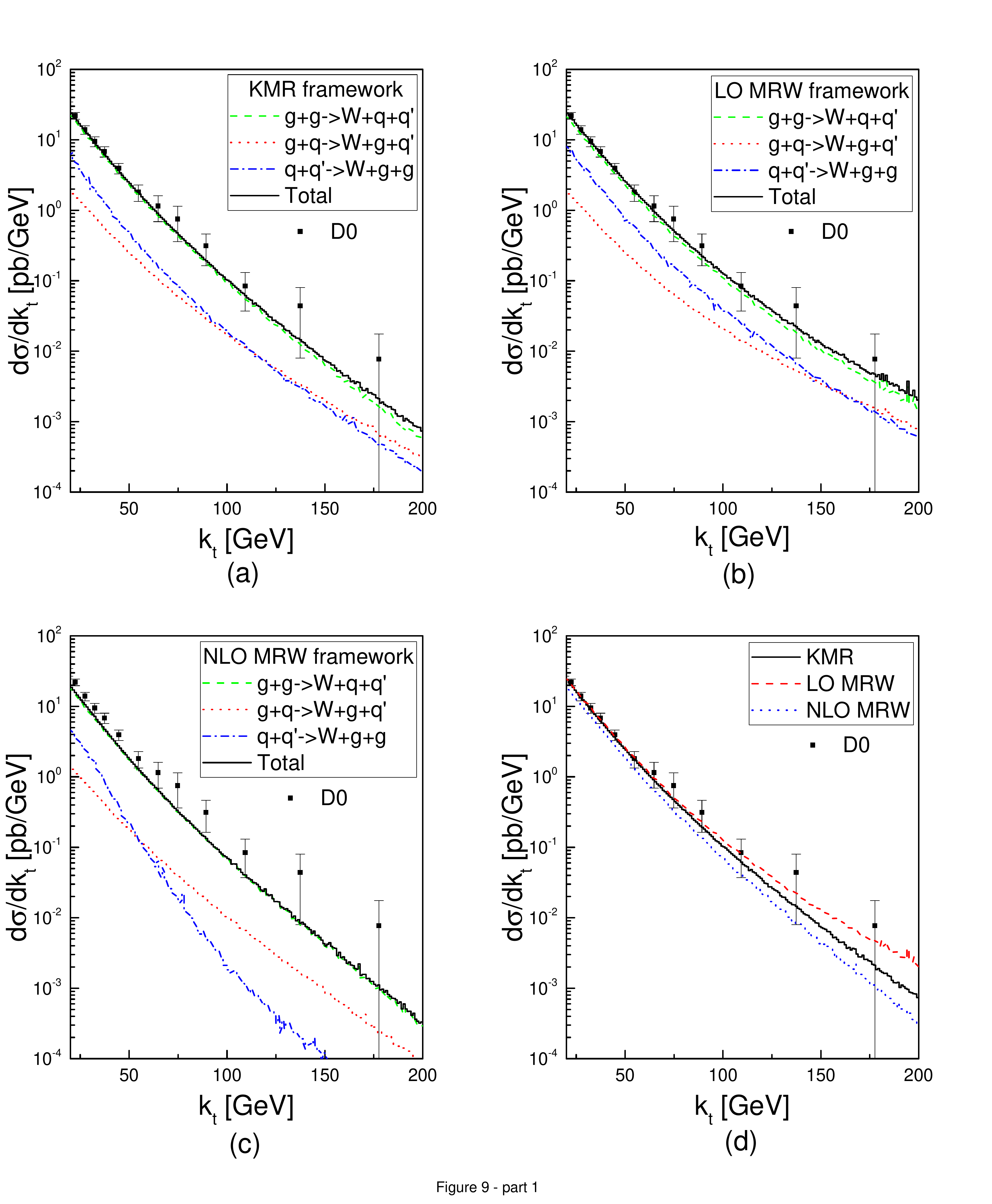}
\includegraphics[scale=0.2]{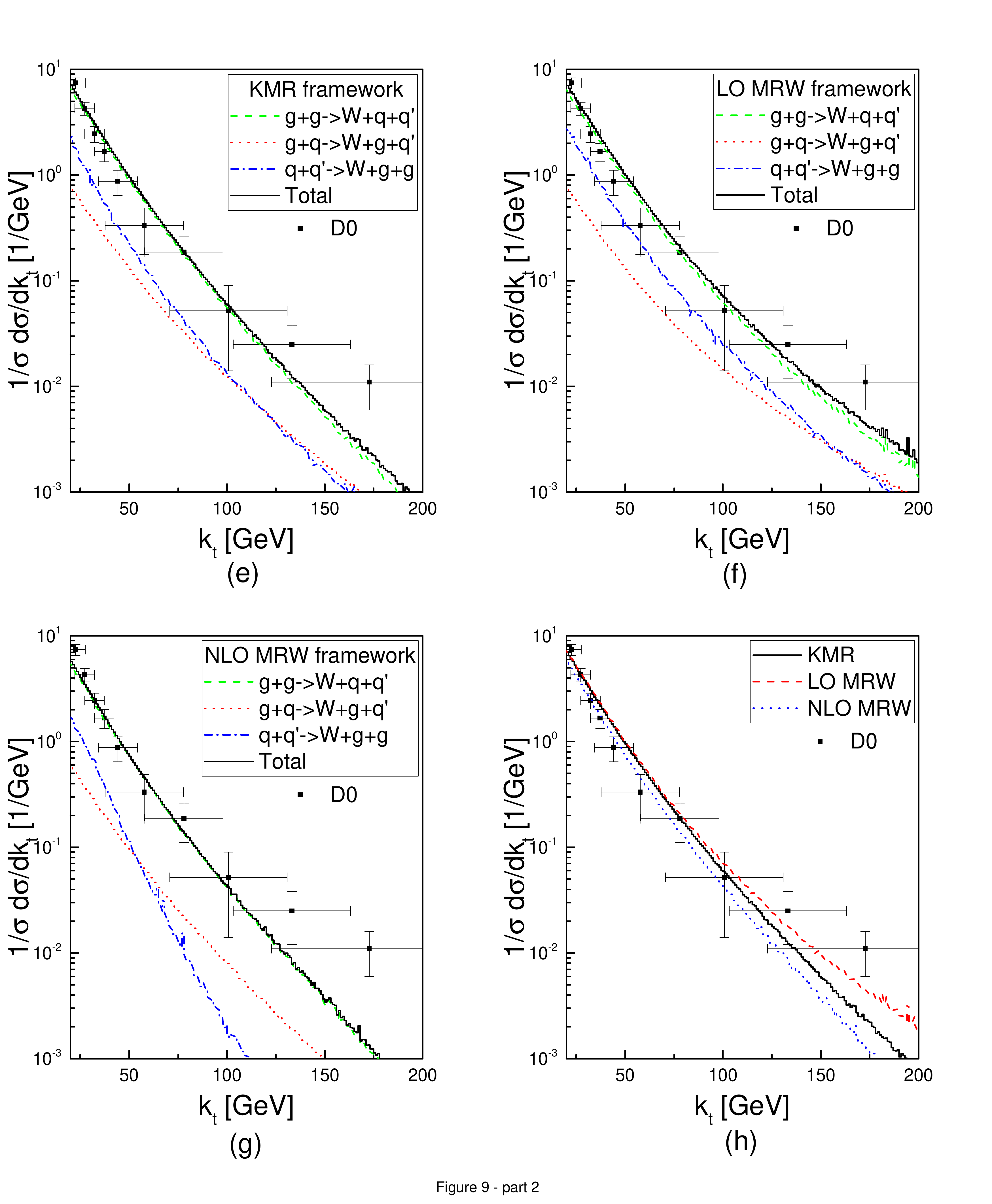}
\caption{The production rate of the $W^{\pm}$ boson in $E_{CM}=1.8
\; TeV$. The labels (a), (b) and (c) compare the contributions of
the individual sub-processes in their respective frameworks. The
total values of differential cross-section in these frameworks are
subjected to a comparison with the data of the $D0$ collaboration
\cite{D02001} separately, in the label (d). This very same notion is
also presented in the labels (e) through (f), where the $1/\sigma \;
d\sigma/dk_t$ histograms are being compared with each other and with
the data from \cite{D098}.} \label{fig9}
\end{figure}

\begin{figure}[ht]
\centering
\includegraphics[scale=0.2]{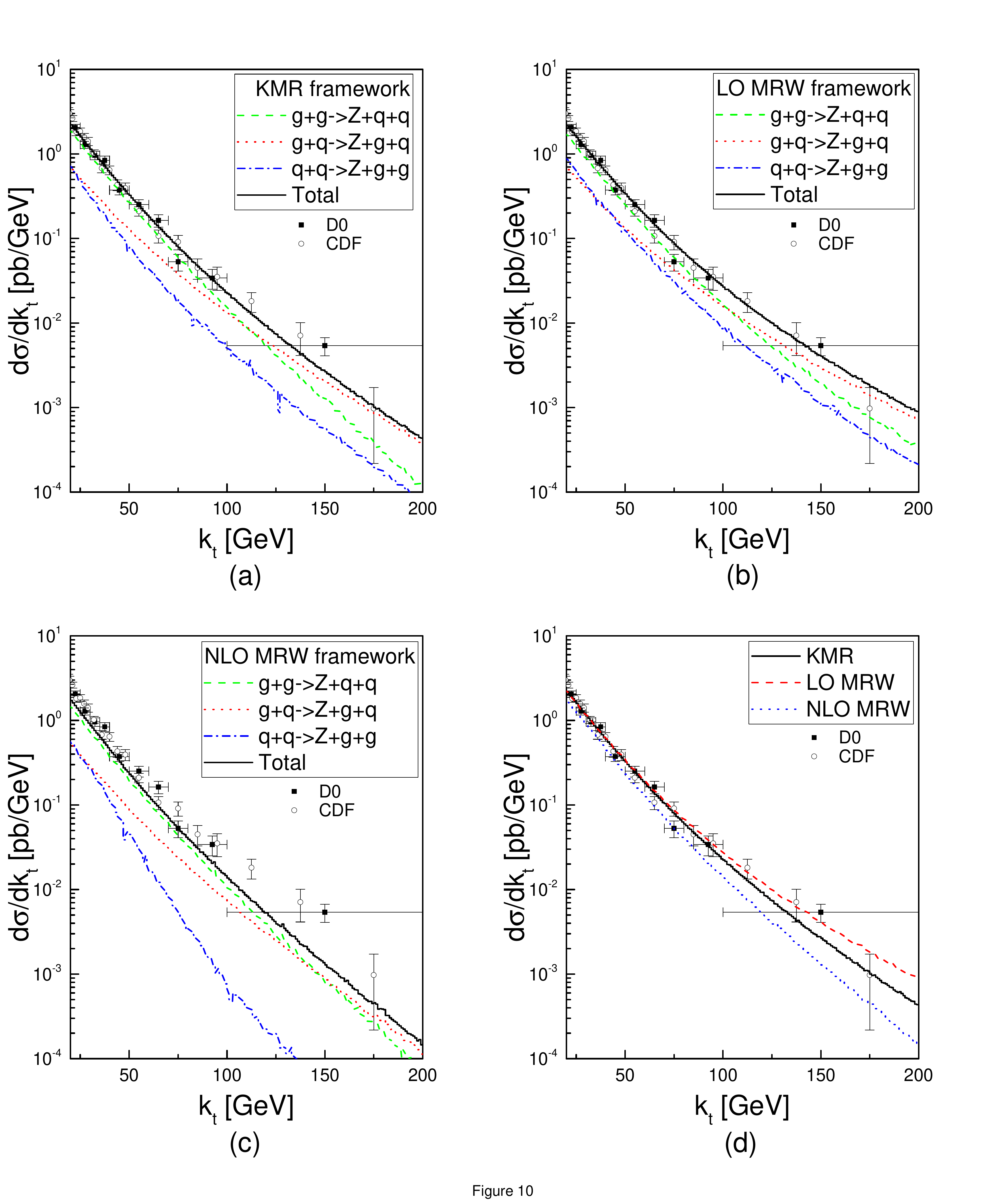}
\caption{The production rate of the $Z^{0}$ boson in $E_{CM}=1.8 \;
TeV$. The notions of the diagrams are the same as in the Fig.
\ref{fig9}.} \label{fig10}
\end{figure}

\begin{figure}[ht]
\centering
\includegraphics[scale=0.2]{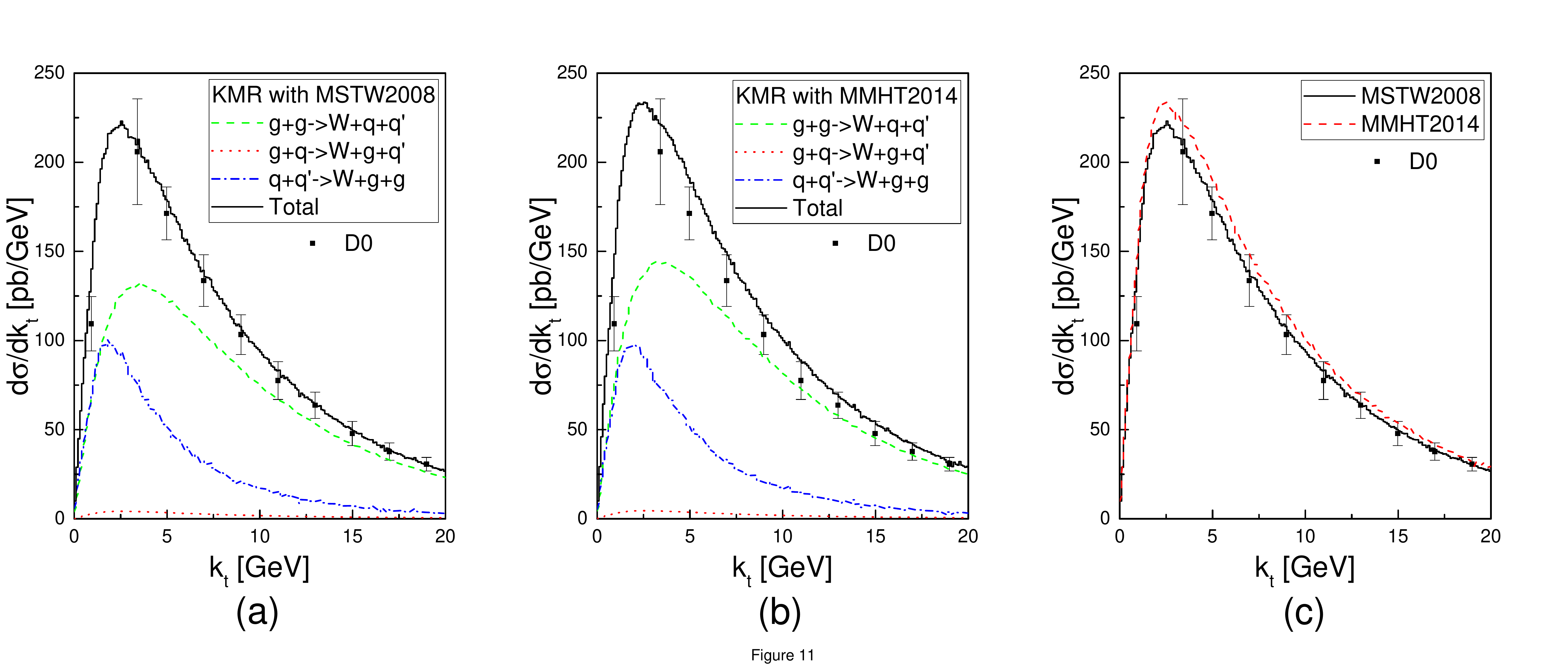}
\caption{Comparison of the differential cross-section of the
$W^{\pm}$ production, using the $UPDF$ of $KMR$, prepared with the
$PDF$ of $MSTW2008$ (label (a)) and $MMHT2014$ (label (b)). label
(c) shows their difference relative to the experimental data of the
$D0$ collaboration, reference \cite{D02001}.} \label{fig11}
\end{figure}

\begin{figure}[ht]
\centering
\includegraphics[scale=0.2]{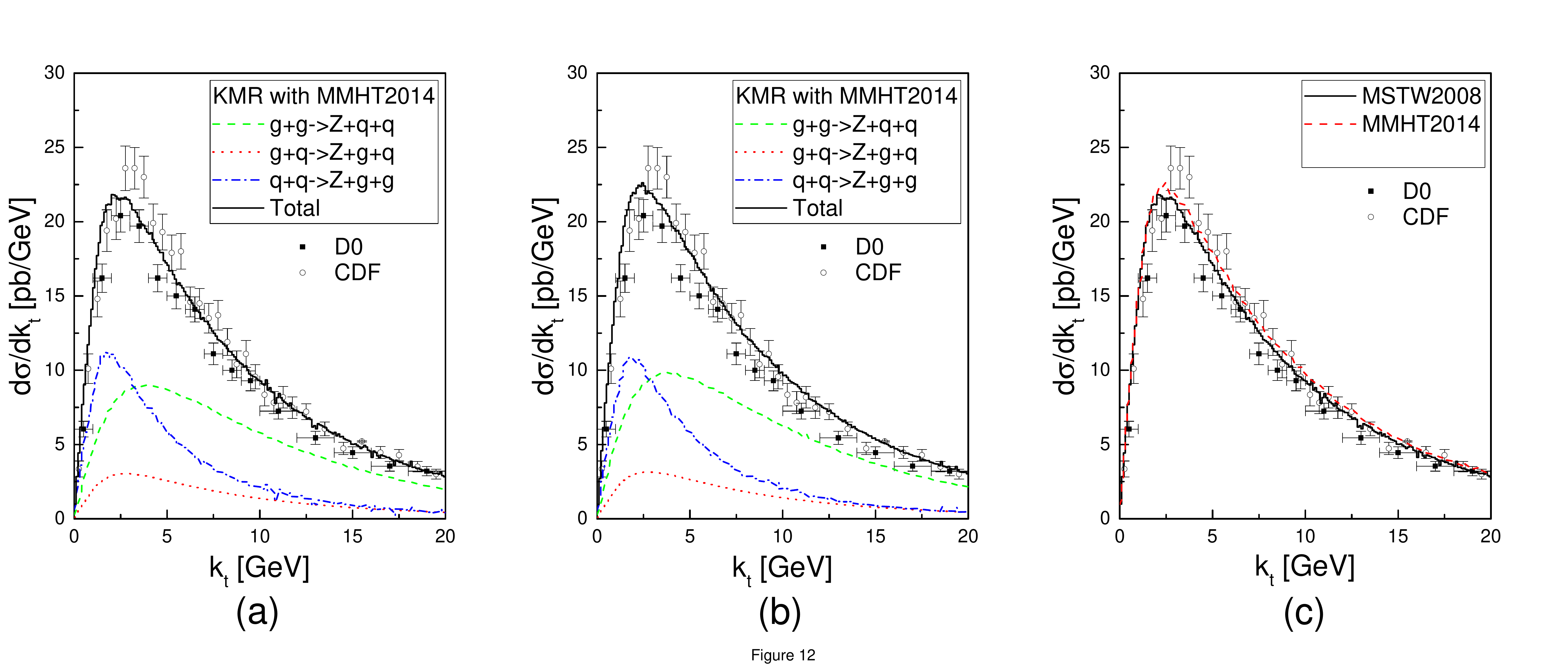}
\caption{Comparison of the differential cross-section of the $Z^{0}$
production, using the $UPDF$ of $KMR$, prepared with the $PDF$ of
$MSTW2008$ (label (a)) and $MMHT2014$ (label (b)). label (c) shows
their difference relative to the experimental data of the $D0$ and
$CDF$ collaborations, references \cite{D098, CDF2000}.}
\label{fig12}
\end{figure}

\begin{figure}[ht]
\centering
\includegraphics[scale=0.2]{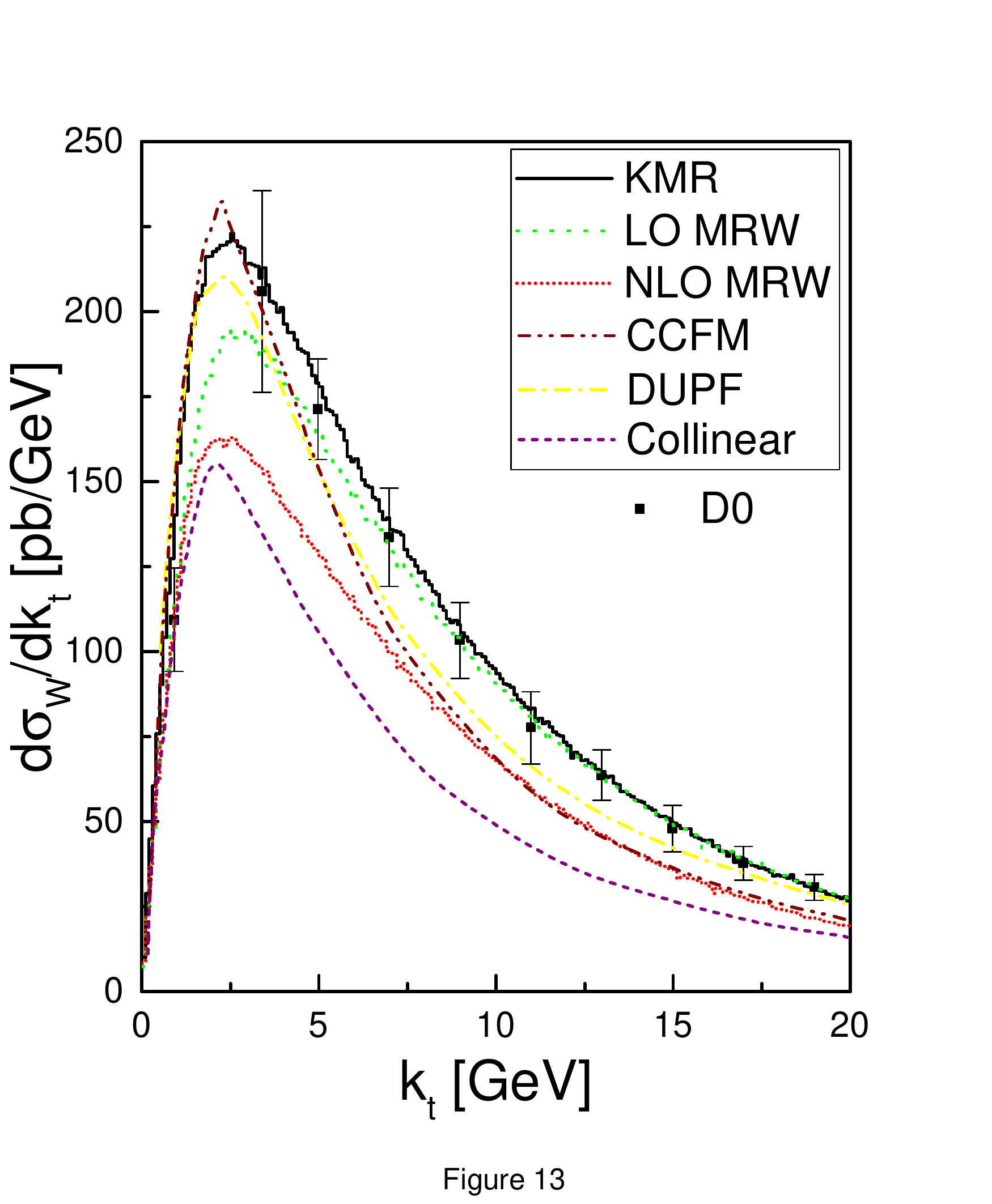}
\caption{The differential cross-section of the production of the
$W^{\pm}$, calculated in different frameworks, against the
transverse momentum of the produced gauge boson at $E_{CM} =
1.8\;TeV$. The notions of the histograms are as follows: the
continues black histogram represents the calculation in using the
$KMR$ $UPDF$, the dotted green histogram is prepared in $LO$ $MRW$
framework and the short-dotted red in the $NLO$ $MRW$. To perform
these calculations, we have utilized the $PDF$ of $MSTW2008$. the
brown dot-dot-dashed histogram in produced using the $CCFM$
$TMD\;PDF$ (reference \cite{LIP1}). The yellow dotted-dashed
histogram is calculated, utilizing the doubly unintegrated parton
distributions (DUPDF) in the framework of $(k_t-z)$-factorization,
reference \cite{WattWZ}. The purple short-dashed histogram is
calculated in the collinear framework.} \label{fig13}
\end{figure}

\begin{figure}[ht]
\centering
\includegraphics[scale=0.2]{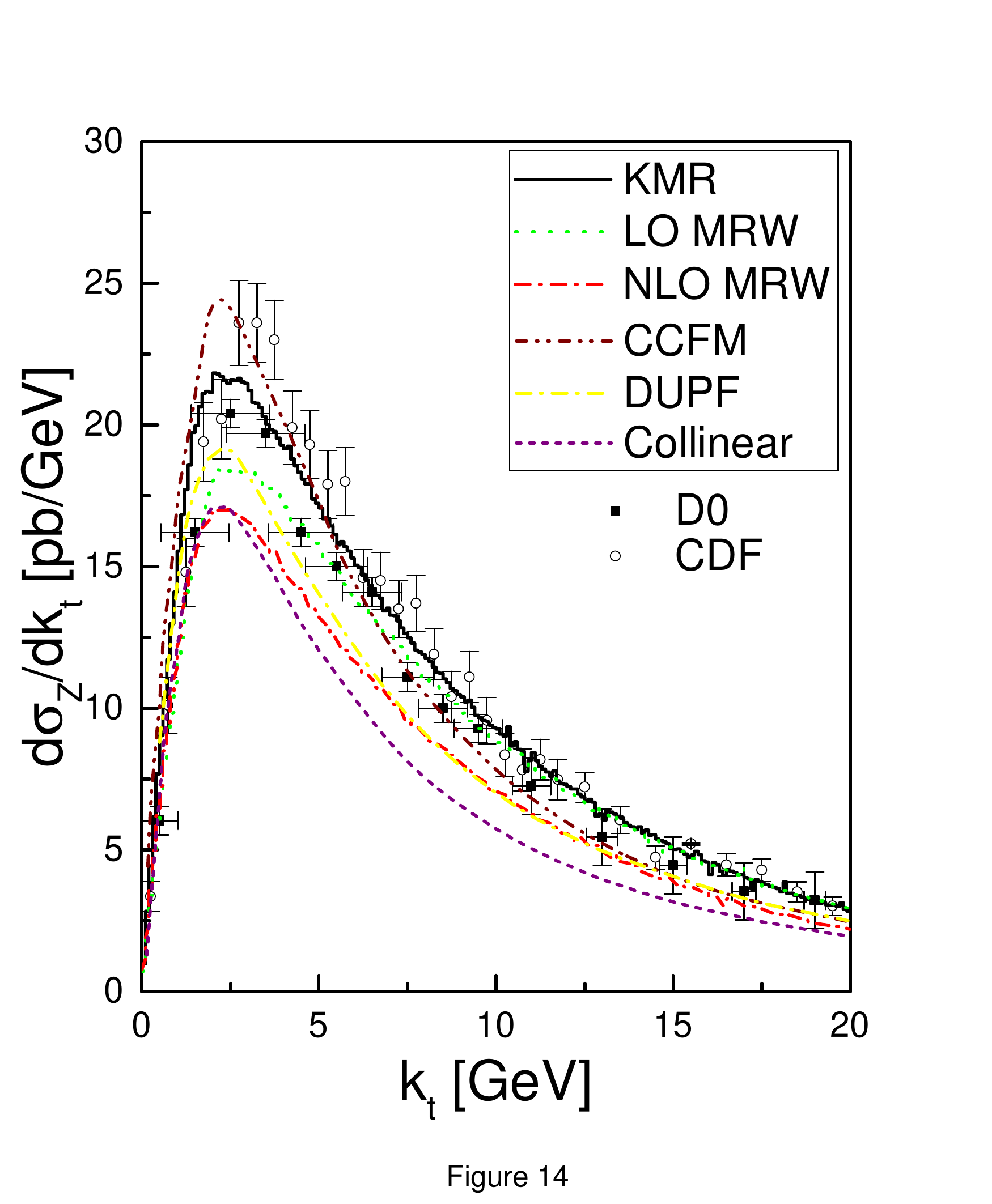}
\caption{The differential cross-section of the production of the
$Z^{0}$, calculated in different frameworks, against the transverse
momentum of the produced gauge boson at $E_{CM} = 1.8\;TeV$. The
notions of the histograms are the same as in Fig.\ref{fig13}.}
\label{fig14}
\end{figure}

\begin{figure}[ht]
\centering
\includegraphics[scale=0.2]{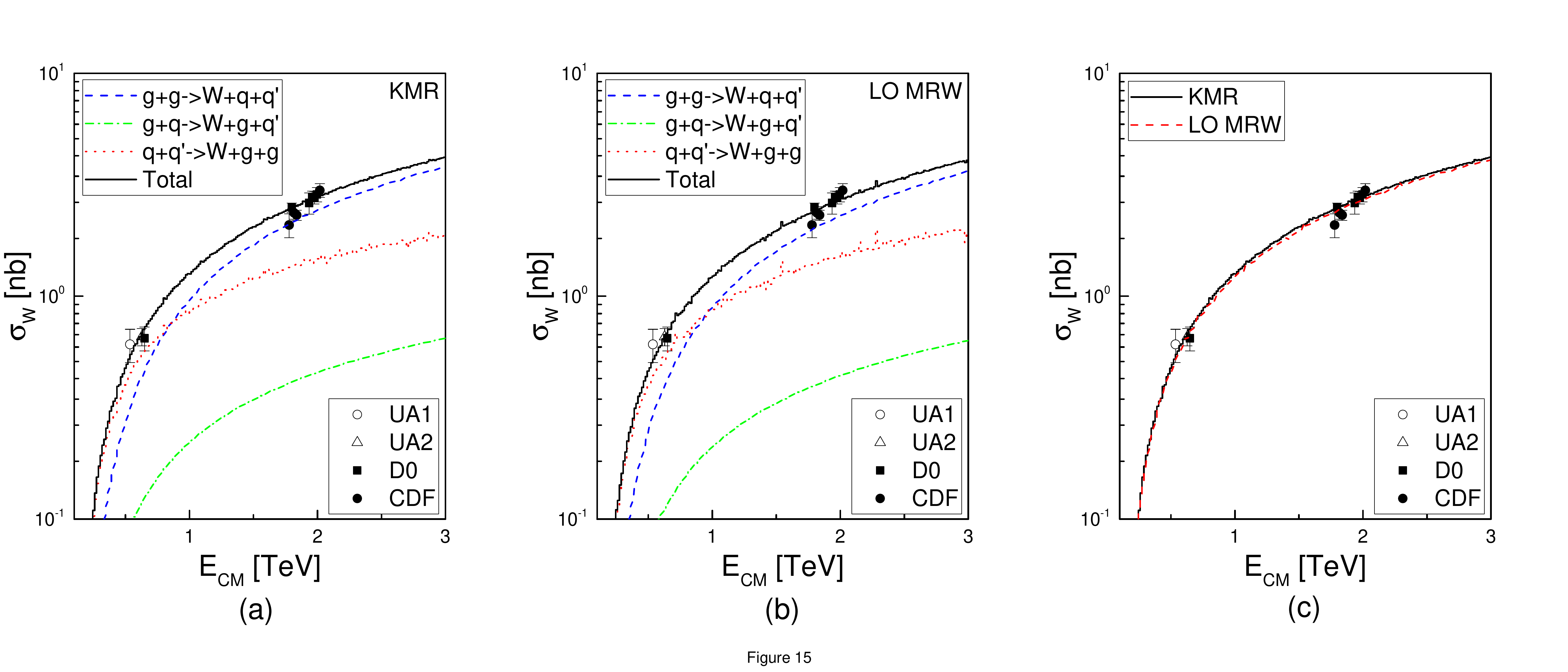}
\caption{The cross-section of the production of the $W^{\pm}$ bosons
as a function of the center-of-mass energy, $E_{CM}$. The
experimental data are acquired from the $UA1$, $UA2$, $D0$ and $CDF$
collaborations, references \cite{UA1, UA2, CDF96, CDF2000, D095,
D098, D02000-1, D02000-2, D02001}. The calculations are performed
using the $KMR$ and the $LO \; MRW$ $UPDF$. We have omitted the
$NLO$ $UPDF$ results here, to save computation data.} \label{fig15}
\end{figure}

\begin{figure}[ht]
\centering
\includegraphics[scale=0.2]{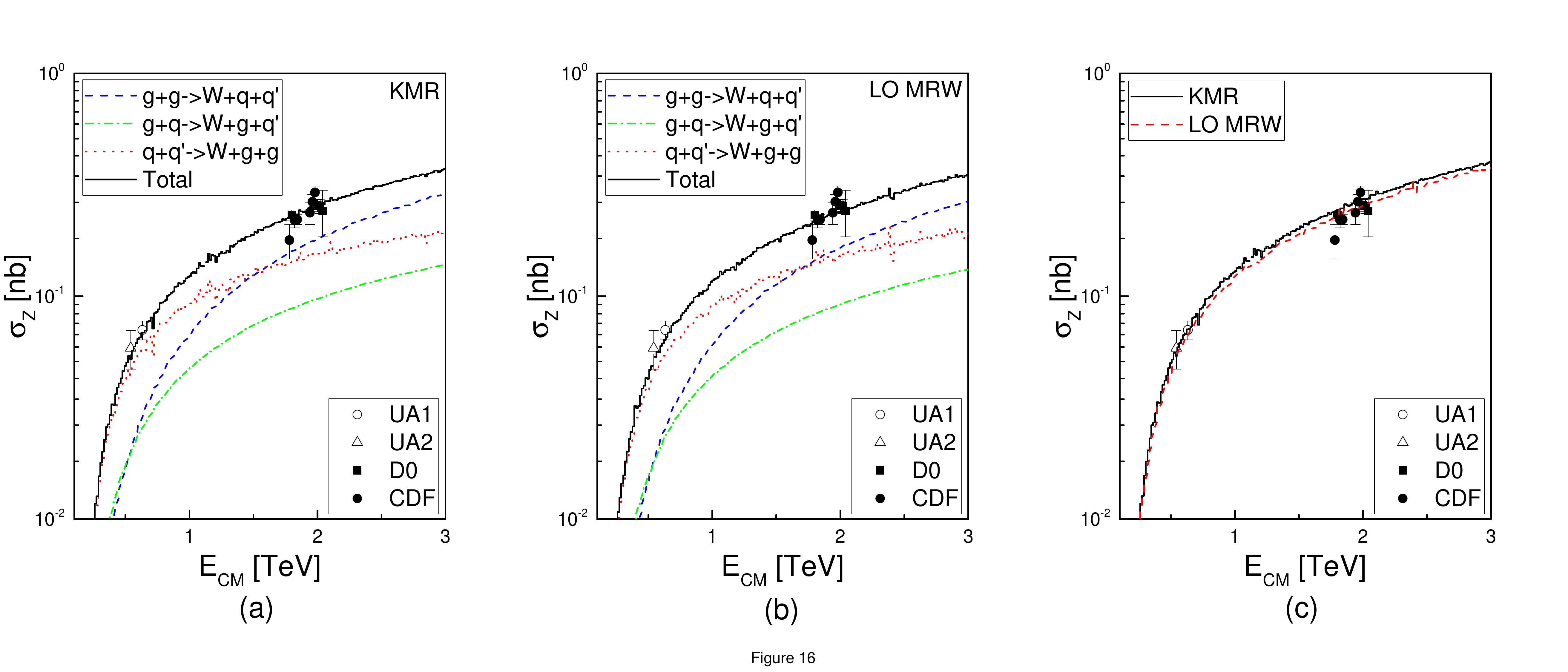}
\caption{The cross-section of the production of the $Z^{0}$ bosons
as a function of the center-of-mass energy, $E_{CM}$. The notation
of the diagram is the same as in the figure \ref{fig15}.}
\label{fig16}
\end{figure}

\begin{figure}[!ht]
\centering
\includegraphics[scale=0.2]{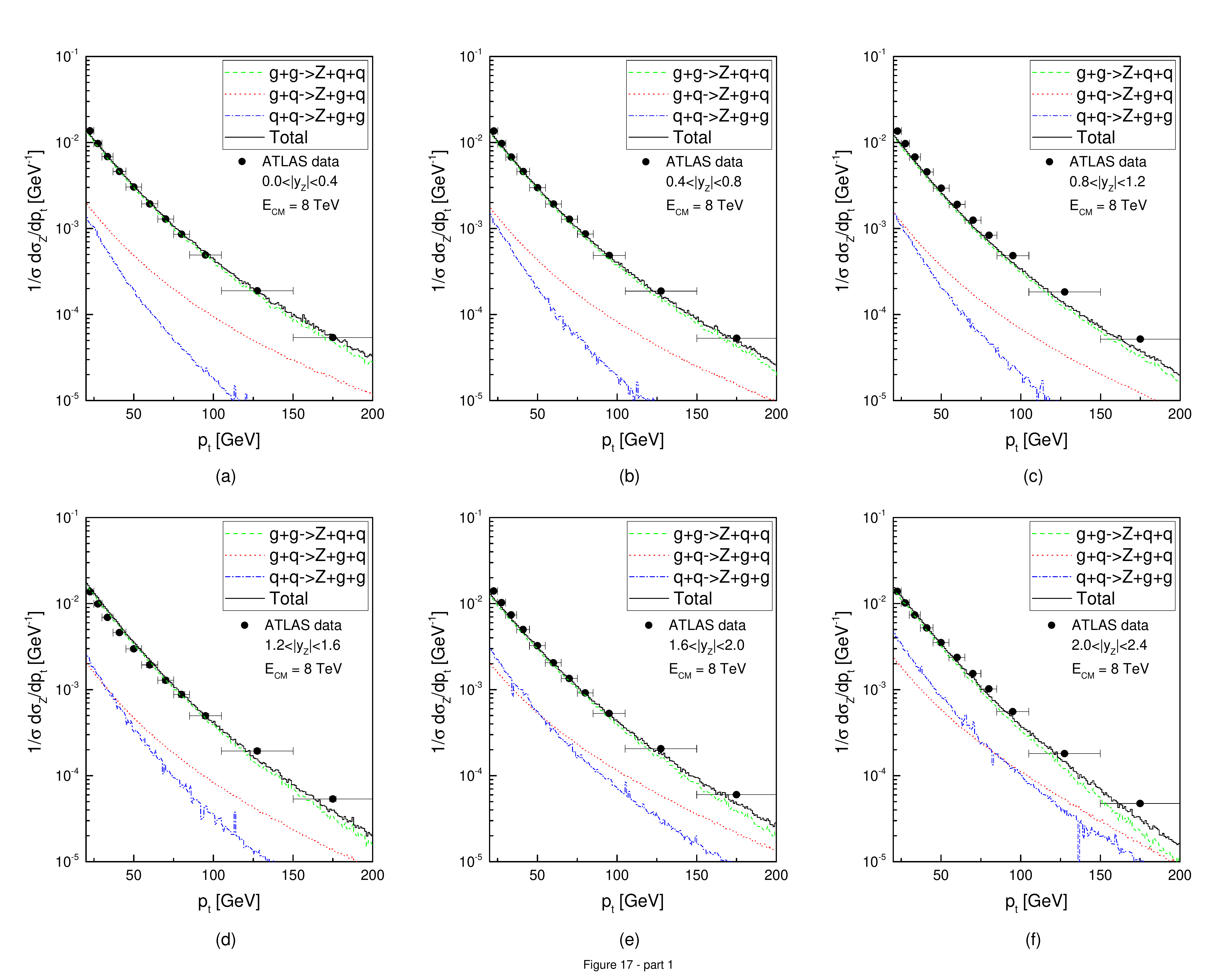}
\includegraphics[scale=0.2]{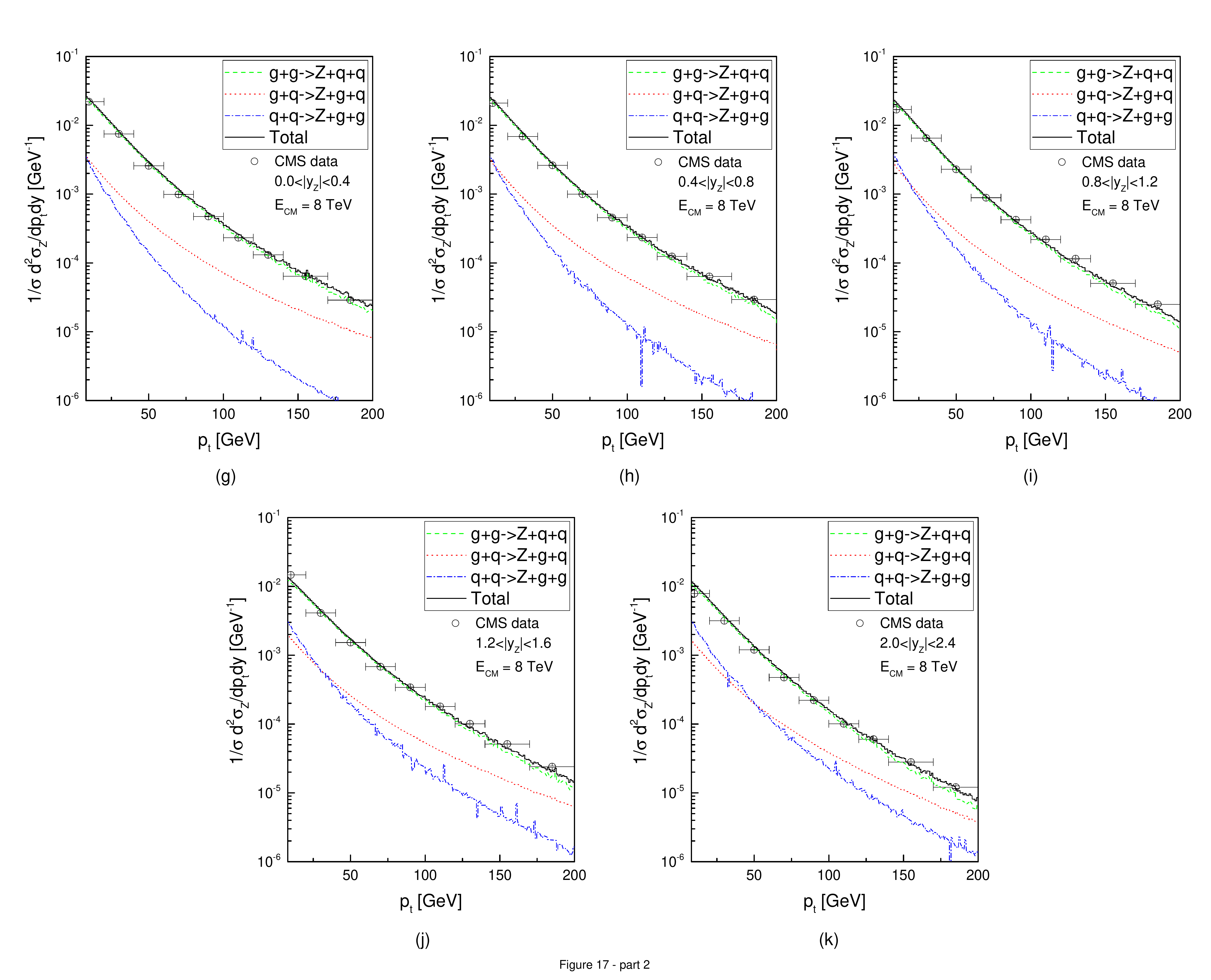}
 \caption{Production of the $Z^{0}$ boson in $E_{CM}=8 \;
TeV$, using the $KMR$ approach. The individual contributions from
the partonic sub-processes are presented and the total values of
(single and double) differential cross-sections are subjected to
comparison with the data of the $ATLAS$ (black circles) and $CMS$
(white circles) collaborations \cite{ATLAS2016,CMS2015}. The labels
(a) through (f) illustrate the results of our calculations for
single differential cross-section of the production of $Z^0$, in the
given rapidity regions. The results for double differential
cross-section are presented in the this figure with labels (g)
through (h).} \label{fig17}
\end{figure}

Given that we are interested in the calculation of the matrix
element squared for each process, one immediately concludes that the
$|\mathcal{M}^{gg}|^2 = |\mathcal{M}^{qq}|^2$. Therefore it is
sufficient to calculate the invariant amplitudes for the Feynman
diagrams of the figure \ref{fig2} , the panels (b) and (c), which
can be written as follows:
\begin{equation}
\mathcal{M}^{ab}= \sum_{i=1}^8 \mathcal{M}_i^{ab}, \;\;\; a,b=q,g,
\end{equation}
with
$$
\mathcal{M}_1^{qg} = g^2_s \; u(k_1) \; t^a \gamma_{\mu}
\epsilon^{\mu}(p_1) \; { \cancel{(k_1 - p_1)} + m \over (k_1 -
p_1)^2 - m^2} \; G^{\lambda}_{W,Z} \epsilon_{\lambda}(p_3) \;
$$
\begin{equation}
{ \cancel{(k_2 + p_2)} + m \over (k_2 + p_2)^2 - m^2} \; t^b
\gamma_{\nu} \epsilon^{\nu}(k_2) \; \bar{u}(p_2),
\end{equation}
$$
\mathcal{M}_2^{qg} = g^2_s \; u(k_1) \; t^b \gamma_{\nu}
\epsilon^{\nu}(k_2) \; { \cancel{(k_1 + k_2)} + m \over (k_1 +
k_2)^2 - m^2} \; G^{\lambda}_{W,Z} \epsilon_{\lambda}(p_3) \;
$$
\begin{equation}
{ \cancel{(k_2 + k_2 - p_3)} + m \over (k_2 + k_2 - p_3)^2 - m^2} \;
t^a \gamma_{\mu} \epsilon^{\mu}(p_1) \; \bar{u}(p_2),
\end{equation}
$$
\mathcal{M}_3^{qg} = 2 g^2_s \; u(k_1) \; t^a \gamma_{\mu}
\epsilon^{\mu}(p_1) \; { \cancel{(k_1 - p_1)} + m \over (k_1 -
p_1)^2 - m^2} \;
$$
\begin{equation}
t^b \gamma_{\nu} \epsilon^{\nu}(k_2) \; { \cancel{(p_2 + p_3)} + m
\over (p_2 + p_3)^2 - m^2} \; G^{\lambda}_{W,Z}
\epsilon_{\lambda}(p_3) \; \bar{u}(p_2),
\end{equation}
$$
\mathcal{M}_4^{qg} = 2 g^2_s \; u(k_1) \; t^a \gamma_{\mu}
\epsilon^{\mu}(k_2) \; { \cancel{(k_1 + k_2)} + m \over (k_1 +
k_2)^2 - m^2} \;
$$
\begin{equation}
t^b \gamma_{\nu} \epsilon^{\nu}(p_1) \; { \cancel{(k_1 + k_2 - p_1)}
+ m \over (k_1 + k_2 - p_1)^2 - m^2} \; G^{\lambda}_{W,Z}
\epsilon_{\lambda}(p_3) \; \bar{u}(p_2),
\end{equation}
$$
\mathcal{M}_5^{qg} = g^2_s \; u(k_1) \; \gamma^{\rho} C^{\mu \nu
\rho}(k_2,-p_1,p_1 - k_2) { \epsilon_{\mu} \epsilon_{\nu} \over (k_2
- p_1)^2} f^{abc} t^c\;
$$
\begin{equation}
{ \cancel{(p_2 + p_3)} + m \over (p_2 + p_3)^2 - m^2} \;
G^{\lambda}_{W,Z} \epsilon_{\lambda}(p_3) \; \bar{u}(p_2),
\end{equation}
$$
\mathcal{M}_6^{qg} = g^2_s \; u(k_1) \; G^{\lambda}_{W,Z}
\epsilon_{\lambda}(p_3) \; { \cancel{(k_1 - p_3)} + m \over (k_1 -
p_3)^2 - m^2} \;
$$
\begin{equation}
\gamma^{\rho} C^{\mu \nu \rho}(k_2,-p_1,p_1 - k_2) { \epsilon_{\mu}
\epsilon_{\nu} \over (k_2 - p_1)^2} f^{abc} t^c\; \bar{u}(p_2),
\end{equation}
and
$$
\mathcal{M}_1^{gg} = g^2_s \; \bar{u}(p_1) \; t^a \gamma_{\mu}
\epsilon^{\mu}(k_1) \; { \cancel{(p_1 - k_1)} + m \over (p_1 -
k_1)^2 - m^2} \; G^{\lambda}_{W,Z} \epsilon_{\lambda}(p_3) \;
$$
\begin{equation}
{ \cancel{(p_2 + k_2)} + m \over (p_2 + k_2)^2 - m^2} \; t^b
\gamma_{\nu} \epsilon^{\nu}(k_2) \; u(p_2),
\end{equation}
$$
\mathcal{M}_2^{gg} = g^2_s \; \bar{u}(p_1) \; t^a \gamma_{\mu}
\epsilon^{\mu}(k_1) \; { \cancel{(k_2 - p_1)} + m \over (k_2 -
p_1)^2 - m^2}  \; G^{\lambda}_{W,Z} \epsilon_{\lambda}(p_3) \;
$$
\begin{equation}
{ \cancel{(k_1 + p_2)} + m \over (k_1 + p_2)^2 - m^2} \; t^b
\gamma_{\nu} \epsilon^{\nu}(k_2) \; u(p_2),
\end{equation}
$$
\mathcal{M}_3^{gg} = g^2_s \; \bar{u}(p_1) { \cancel{(p_1 + p_3)} +
m \over (p_1 + p_3)^2 - m^2}  \; t^a \gamma_{\mu}
\epsilon^{\mu}(k_1) \;
$$
\begin{equation}
{ \cancel{(p_1 + p_3 - k_1)} + m \over (p_1 + p_3 - k_1)^2 - m^2} \;
G^{\lambda}_{W,Z} \epsilon_{\lambda}(p_3) \; t^b \gamma_{\nu}
\epsilon^{\nu}(k_2) \; u(p_2),
\end{equation}
$$
\mathcal{M}_4^{gg} = g^2_s \; \bar{u}(p_1) \; G^{\lambda}_{W,Z}
\epsilon_{\lambda}(p_3) \; t^a \gamma_{\mu} \epsilon^{\mu}(k_1) \; {
\cancel{(p_1 - k_1)} + m \over (p_1 - k_1)^2 - m^2} \;
$$
\begin{equation}
t^b \gamma_{\nu} \epsilon^{\nu}(k_2) \; { \cancel{(p_1 - k_1 - k_2)}
+ m \over (p_1 - k_1 - k_2)^2 - m^2} \; u(p_2),
\end{equation}
$$
\mathcal{M}_5^{gg} = g^2_s \; \bar{u}(p_1) \; G^{\lambda}_{W,Z}
\epsilon_{\lambda}(p_3) \; { \cancel{(p_1 + p_3)} + m \over (p_1 +
p_3)^2 - m^2}  \; t^b \gamma_{\nu} \epsilon^{\nu}(k_2) \;
$$
\begin{equation}
{ \cancel{(p_1 + p_3 - k_2)} + m \over (p_1 + p_3 - k_2)^2 - m^2} \;
t^a \gamma_{\mu} \epsilon^{\mu}(k_1) \; u(p_2),
\end{equation}
$$
\mathcal{M}_6^{gg} = g^2_s \; \bar{u}(p_1) \; G^{\lambda}_{W,Z}
\epsilon_{\lambda}(p_3) \; t^b \gamma_{\nu} \epsilon^{\nu}(k_2) \; {
\cancel{(p_1 - k_2)} + m \over (p_1 - k_2)^2 - m^2} \;
$$
\begin{equation}
t^a \gamma_{\mu} \epsilon^{\mu}(k_1) \; { \cancel{(p_1 - k_1 - k_2)}
+ m \over (p_1 - k_1 - k_2)^2 - m^2} \; u(p_2),
\end{equation}
$$
\mathcal{M}_7^{gg} = g^2_s \; \bar{u}(p_1) \; \gamma^{\rho} C^{\mu
\nu \rho}(k_1,k_2,-k_1 - k_2) { \epsilon_{\mu} \epsilon_{\nu} \over
(k_1 + k_2)^2} f^{a b c} t^c
$$
\begin{equation}
{ \cancel{(p_1 - k_1 - k_2)} + m \over (p_1 - k_1 - k_2)^2 - m^2} \;
G^{\lambda}_{W,Z} \epsilon_{\lambda}(p_3) \; u(p_2),
\end{equation}
$$
\mathcal{M}_8^{gg} = g^2_s \; \bar{u}(p_1) \; G^{\lambda}_{W,Z}
\epsilon_{\lambda}(p_3) \; { \cancel{(p_1 - p_3)} + m \over (p_1 -
p_3)^2 - m^2} \;
$$
\begin{equation}
\gamma^{\rho} C^{\mu \nu \rho}(k_1,k_2,-k_1 - k_2) { \epsilon_{\mu}
\epsilon_{\nu} \over (k_1 + k_2)^2} f^{a b c} t^c \; u(p_2),
\end{equation}
where $g_s$ is the running coupling constant for $QCD$ and
$G^{\lambda}_{W,Z}$ represents the vertex of the electro-weak gauge
vector bosons with quarks:
$$
G^{\lambda}_{W} = {e_{em} \over 2\sqrt{2} sin\theta_w}
\gamma^{\lambda} (1-\gamma^5) V_{qq'}
$$
\begin{equation}
G^{\lambda}_{Z} = {e_{em} \over sin 2\theta_w} \gamma^{\lambda}
\left[ I_{3,q}(1-\gamma^5) -2 e_q sin^2 \theta_w \right].
\end{equation}
$\theta_w$ is the Weinberg angle, $V_{qq'}$ is the corresponding
$CKM$ matrix element and $I_{3,q}$ is the weak isospin component of
the quark $q$. Additionally, the standard $QCD$ three-gluon coupling
can be written as follows:
\begin{equation}
C^{\mu \nu \rho}(k_1,k_2,k_3) = g^{\mu \nu} (k_2 - k_1)^{\rho} +
g^{\nu \rho} (k_3 - k_2)^{\mu} + g^{\rho \mu} (k_1 - k_3)^{\nu}.
\end{equation}
With the above information, one has enough tools to calculate the
matrix elements of the equation (\ref{eq25}).

\end{document}